\documentclass[twocolumn,aps,prd,preprintnumbers,superscriptaddress,nofootinbib,10pt]{revtex4-1}
\usepackage{graphicx}
\usepackage{amsmath,amssymb}
\usepackage{amsfonts}
\usepackage{slashed}
\usepackage{dcolumn}

\usepackage{colordvi}
\usepackage{color}
\usepackage{xcolor}

\newcommand{\Dlr}{\overset{\leftrightarrow}{D}}
\newcommand{\Dl}{\overset{\leftarrow}{D}}
\newcommand{\Dr}{\overset{\rightarrow}{D}}
\newcommand{\cO}{\mathcal {O}}
\newcommand{\dummy}{\Box}

\begin{document}

\title{Lattice results for the longitudinal spin structure and color forces
  on quarks in a nucleon}

\author{S.~B\"urger}
\author{T.~Wurm}
\author{M.~L\"offler}
\author{M.~G\"ockeler}
\author{G.~Bali}
\author{S.~Collins}
\author{A.~Sch\"afer}
\affiliation{University of Regensburg, 93040 Regensburg, Germany}
\author{A.~Sternbeck}
\affiliation{Friedrich Schiller University Jena, Max-Wien-Platz 1, 07743 Jena,
  Germany}

\collaboration{RQCD Collaboration}

\begin{abstract}
Using lattice QCD, we calculate the twist-2 contribution $a_2$ to the third
Mellin moment of the spin structure functions $g_1$ and $g_2$ in the nucleon.
In addition we evaluate the twist-3 contribution $d_2$. Our computations
make use of $N_f=2+1$ gauge field ensembles generated by the
Coordinated Lattice Simulations (CLS) effort. Neglecting quark-line
disconnected contributions we obtain as our best estimates
$a_2^{(p)}= 0.069(17)$, $d_2^{(p)}= 0.0105(68)$ and
$a_2^{(n)}= 0.0068(88)$, $d_2^{(n)}= -0.0009(70)$ for the proton
and the neutron, respectively, where we use the normalizations given in
Eqs.~\eqref{eq:a2def} and \eqref{eq:d2def}.
While the $a_2$ results have been converted to the
$\overline{\mathrm{MS}}$ scheme using three-loop perturbation theory,
the numbers for $d_2$ are given in the regularization independent momentum
subtraction (RI$^\prime$-MOM) scheme, i.e.,   
the conversion has been performed only in tree-level perturbation theory.
The $d_2$ results can be interpreted as
corresponding to a transverse color Lorentz force on a quark in a
transversely polarized proton of size $F^{(u)} = 116(61) \, \mbox{MeV/fm}$
and $F^{(d)} = -38(66) \, \mbox{MeV/fm}$ for $u$ and $d$ quarks,
respectively. The error estimates quoted include statistical and
systematic uncertainties added in quadrature.
\end{abstract}

\maketitle

\section{Introduction}
\label{sec:intro}

For a number of reasons hadron spin structure has attracted intense
interest for more than two decades with no sign of attenuation. Quite
to the contrary, CEBAF@12GeV~\cite{Burkert:2018nvj} and the
EIC~\cite{AbdulKhalek:2021gbh} promise to bring such
investigations to a higher level both with respect to the precision and
the variety of observables investigated. This provides a strong motivation
also to update theory predictions for the relevant spin-dependent
quantities. The central goal of the JLAB and BNL programs
is to better understand the structure of hadrons. This includes
multiparton correlations, which are parametrized by higher-twist coefficients.

The most prominent such example is the matrix element $d_2$, the third
Mellin moment ($x^2$) of the twist-3 contribution to the helicity
structure function $g_2(x,Q^2)$ of deep-inelastic longitudinally polarized 
lepton-nucleon scattering. As this corresponds to the
lowest-dimensional nontrivial chiral-even twist-3 matrix element,
$d_2$ is of particular theoretical and phenomenological interest.
For instance, the same correlations of quarks and gluons constitute
the leading contribution to the Qiu-Sterman distributions~\cite{Qiu:1991pp},
which play a central role in the collinear factorization of single
spin asymmetries. These distributions represent the limit of vanishing
impact parameter $\vec{b}_\perp=\vec{0}$
of the Sivers functions~\cite{Sivers:1989cc}, i.e., of the
transverse-momentum dependent parton distribution functions
$f_{1T}^{\perp}(x,\vec{b}_\perp,Q^2)$ that describe the distribution of an
unpolarized quark inside a transversely polarized nucleon.
The measurement of the Sivers distributions in polarized semi-inclusive
deep-inelastic scattering and in Drell-Yan experiments is one of the
main goals of the experimental programs at JLAB and the EIC.
For more details, see, e.g., Refs.~\cite{Scimemi:2019gge,Bury:2020vhj}.

Neglecting the twist-3 contributions, $g_2$ can be obtained from 
the helicity structure function $g_1(x,Q^2)$. This involves invoking
the well-known Wandzura-Wilczek relation~\cite{Wandzura:1977qf,Jaffe:1990qh}.
However, a remarkable property of $g_2$ is that the twist-3 contribution is not
power suppressed in $1/Q$, relative to its twist-2 part. Nevertheless,
its determination from longitudinally polarized deep-inelastic scattering
experiments alone still represents a serious challenge and new
high precision measurements are planned at JLAB and the EIC.
In order to match the expected statistical precision of the planned
experiments, a much improved theoretical understanding of higher-twist
contributions is needed and $d_2$ is the ideal starting point.

The matrix element $d_2$ has another very interesting
phenomenological interpretation: As was argued in
Refs.~\cite{Burkardt:2008ps,Aslan:2019jis} it is related to the average
transverse color Lorentz force acting on a quark $q$ in a nucleon
which moves in the $z$ direction and is transversely polarized.
More explicitly, the $y$-component of the color Lorentz force is given by
\begin{equation} \label{eq:force}
\begin{split}
F^{q,y}(0) &= -\frac{1}{\sqrt{2} p^+}
  \langle p,s|\bar{\psi}_q (0) \gamma^+gG^{+y}(0) \psi_q(0)|p,s \rangle \\
  &= \sqrt{2} p^+ s^x d_2^{(q)} \,.
\end{split}
\end{equation}
Here the four-momentum $p$ of the nucleon state $| p,s \rangle$ has been
chosen as $(p^\mu) = p^+(1,0,0,1)/\sqrt{2}$, the spin vector $s$ is
normalized according to $s^2 = - m_N^2$, $G^{\mu \nu}$ denotes the color
field strength tensor and $g$ is the strong coupling constant.
Using  the lattice results for the proton presented in
Ref.~\cite{Gockeler:2005vw} (coauthored by some of us)
\begin{equation} \label{eq:2}
\begin{split}  
  d_2^{(u)} &= 0.010 \pm 0.012 \,, \\
  d_2^{(d)} &= -0.0056 \pm 0.0050 \,,
\end{split}  
\end{equation}
estimates for this force were published some time ago
in Ref.~\cite{Burkardt:2008ps}. It appears, however, that these estimates
were affected by a misunderstanding of the respective conventions and
by a sign error noted later in Ref.~\cite{Aslan:2019jis}. The corrected
numbers differ by a factor $- \tfrac{1}{2}$ from those given in 
Ref.~\cite{Burkardt:2008ps} and read
\begin{equation}
\begin{split}  
  F^{u,y}(0) &= 50 \pm 60~{\rm MeV/fm} \,, \\
  F^{d,y}(0) &= - 28 \pm 26~{\rm MeV/fm} \,.
\end{split}  
\end{equation} 

Unfortunately, the errors, which are purely statistical, are very
large. Systematic uncertainties were not estimated, in particular, those
arising from finite lattice spacing. Meanwhile several experiments
have extracted estimates of $d_2^{(p)}$ and
$d_2^{(n)}$~\cite{E154:1997eyc,Abe:1998wq,Anthony:2002hy,Zheng:2004ce,
Airapetian:2011wu,Posik:2014usi,Flay:2016wie,Armstrong:2018xgk},
where the superscript indicates proton or neutron, respectively. These
estimates are found to be quite small compared to various model predictions
but compatible with the old lattice results (considering the large
error bars), see, e.g., Fig.~2 of Ref.~\cite{Armstrong:2018xgk}.
(Actually, the lattice results in this figure should also have been
divided by 2.) We remark that with the natural energy scale for the force
$F$ being $\Lambda_{QCD}^2$ one would not expect $d_2$ to be much smaller
than the central values of this early lattice calculation
given in Eq.~(\ref{eq:2}). So there is hope that with a moderate
reduction of the lattice uncertainties, this time also including
systematics, one may be able to demonstrate that $d_2$ and thus the
average color force $F$ is different from zero.

Let us stress that the experimental and lattice investigations of $d_2^{(p)}$
and $d_2^{(n)}$ are only meant to be the starting point of much
broader investigations. For example, it was also argued in
Ref.~\cite{Aslan:2019jis} that there exists an analogous relationship
between generalized parton distributions 
and force distributions ${\cal F}^i_{\lambda'\lambda}(\vec{b}_\perp)$
in the transverse plane: 
\begin{equation} 
{\cal F}^i_{\lambda'\lambda}(\vec{b}_\perp) =
\int \frac{d^2 \vec{\Delta}_\perp}{(2\pi)^2}
e^{-i \vec{b}_\perp \cdot \vec{\Delta}_\perp}
F^i_{\lambda'\lambda}(\vec{\Delta}_\perp) 
\end{equation}
with 
\begin{equation} 
\begin{split}
F^i_{\lambda'\lambda}(\vec{\Delta}_\perp) &= -\frac{1}{\sqrt{2} p^+}
 \\ & \hspace{-1.4cm} {} \times 
\left \langle p^+,\frac{\vec{\Delta}_\perp}{2},\lambda' \left |\bar{\psi}_q(0)
\gamma^+gG^{+i}(0) \psi_q (0) \vphantom{\frac{{\bf \Delta}_{\perp}}{2}}
\right |p^+, -\frac{\vec{\Delta}_\perp}{2},\lambda \right \rangle \,.
\end{split}
\end{equation}
Here $\lambda$ and $\lambda'$ denote the nucleon polarization and 
$\vec{\Delta}_\perp$ is the transverse momentum conjugate to the
impact parameter $\vec{b}_\perp$.

Another interesting result was derived in the very recent
paper~\cite{Braun:2021aon}, where QCD factorization for quasidistributions
was analyzed up to twist-3. Approaches based on so-called quasi- and
pseudodistribution functions have gained prominence in lattice QCD
calculations of hadron structure observables, due to their prospect
of providing information that goes beyond the computation of Mellin
moments of (generalized) parton distribution
functions, distribution amplitudes etc., see Refs.~\cite{Ji:2013dva,
Cichy:2018mum,Ji:2020ect,Ji:2020brr,Huo:2021rpe,Alexandrou:2020qtt,
Constantinou:2020pek} and references therein.

In Ref.~\cite{Braun:2021aon} it was shown that for quasidistributions
the Wandzura-Wilczek relation~\cite{Wandzura:1977qf,Jaffe:1990qh}
is modified such that twist-2 and twist-3 contributions stay mixed, making
their separate determination on the lattice far more difficult.
Knowing $d_2$ from a direct lattice calculation 
would obviously help to unravel the different contributions.

\section{OPE and renormalization in the continuum}
\label{sec:OPE}

A leading-order OPE (operator product expansion) analysis with massless
quarks shows that the moments of $g_1$ and $g_2$ can be written
as~\cite{Jaffe:1989xx}
\begin{equation} \label{eq:ope-g1}
\begin{split}
2\int_0^1 dx \, &x^n g_1(x,Q^2) \\
&= \frac{1}{2} \sum_q Q_q^2 \,
  E^{(q)}_{1,n}(\mu^2/Q^2,g(\mu)) \, a_n^{(q)}(\mu) \,,
\end{split}
\end{equation}

\begin{equation} \label{eq:ope-g2}
\begin{split}
2\int_0^1 &dx \, x^n g_2(x,Q^2) \\ 
&= \frac{1}{2}\frac{n}{n+1} \sum_q Q_q^2 \,
\big[ E^{(q)}_{2,n}(\mu^2/Q^2,g(\mu)) \, d_n^{(q)}(\mu) \\
&\hspace*{2.5cm} {}- E^{(q)}_{1,n}(\mu^2/Q^2,g(\mu)) \, a_n^{(q)}(\mu) \big] \,,
\end{split}
\end{equation}
where $q$ runs over the light quark flavors with charges $Q_q$
and $\mu$ denotes the renormalization scale.
Equations~\eqref{eq:ope-g1} and \eqref{eq:ope-g2} hold for even $n$,
with $n \ge 0$ for the former and $n \ge 2$ for the latter.
The Wilson coefficients $E_{1,n}^{(q)}$
and $E_{2,n}^{(q)}$ depend on the ratio of scales $\mu^2/Q^2$ and the
running coupling constant $g(\mu)$,
\begin{equation}
E_{i,n}(\mu^2/Q^2,g(\mu)) = 1 + {\mathcal O}(g(\mu)^2) \,.
\label{eq:wilsoncoef}
\end{equation}
To the best of our knowledge, the loop corrections for $E_{2,n}$ 
have not yet been calculated, while they are known up to two-loop order
for $E_{1,n}$~\cite{Zijlstra:1993sh}.
The first-order corrections are flavor independent,
\begin{equation}
E_{1,n}(1,g(\mu)) = 1 + \frac{5}{3} \frac{g(\mu)^2}{16 \pi^2}
  + {\mathcal O}(g(\mu)^4) \,.
\label{eq:wilsoncoef1}
\end{equation}

The reduced matrix elements $a_n^{(q)}(\mu)$ and $d_n^{(q)}(\mu)$ 
are defined as~\cite{Jaffe:1989xx}
\begin{equation} \label{eq:twist2}
\begin{split}
\langle p,s| &\cO^{5 (q)}_{ \{ \sigma\mu_1\cdots\mu_n \} } | p,s \rangle \\
 &= \frac{1}{n+1} a_n^{(q)} [ s_\sigma p_{\mu_1} \cdots p_{\mu_n}
+ \cdots -\mbox{traces}]
\end{split}
\end{equation}

\begin{equation} \label{eq:twist3}
\begin{split}
&\langle p,s| \cO^{5 (q)}_{ [  \sigma \{ \mu_1 ] \mu_2 \cdots \mu_n \} }
  | p,s \rangle \\
&= \frac{1}{n+1}d_n^{(q)} [ (s_\sigma p_{\mu_1} - s_{\mu_1} p_\sigma)
  p_{\mu_2}\cdots p_{\mu_n} + \cdots -\mbox{traces}] 
\end{split}
\end{equation}
in terms of matrix elements of the local operators
\begin{equation} \label{eq:locop}
\cO^{5 (q)}_{\sigma\mu_1\cdots\mu_n}
   = \left(\frac{i}{2}\right)^n \bar{\psi}_q \gamma_{\sigma} \gamma_5
 \Dlr_{\mu_1} \cdots \Dlr_{\mu_n} \psi_q -\mbox{traces} 
\end{equation}
in the nucleon state $| p,s \rangle$.
Here $\Dlr=\Dr-\Dl$ and the symbol $\{\cdots\}$ ($[\cdots]$) indicates
symmetrization (antisymmetrization) of the enclosed indices.
The operator in Eq.~\eqref{eq:twist2} has twist two, whereas the operator
in Eq.~\eqref{eq:twist3} has twist three.
As far as the Wilson coefficients may be considered as flavor independent,
we can define $a_n$ and $d_n$ for the nucleon as
\begin{align} \label{eq:twist2nucleon}
a_n (\mu) &= \sum_q Q_q^2 a_n^{(q)}(\mu) \,, \\  \label{eq:twist3nucleon}
d_n (\mu) &= \sum_q Q_q^2 d_n^{(q)}(\mu)  \,.
\end{align}
Remarkably, in the moments \eqref{eq:ope-g2} of $g_2$ the twist-3 matrix
elements $d_n^{(q)}(\mu)$ are not suppressed relative to the twist-2
matrix elements $a_n^{(q)}(\mu)$.

Note that our definitions of $a_2$ and $d_2$ have been taken from
Ref.~\cite{Jaffe:1989xx}. In many publications alternative definitions
are employed, where $a_2^{\mathrm {alt}} = a_2/2$ and
$d_2^{\mathrm {alt}} = d_2/2$.

Utilizing the equations of motion of massless QCD and the relation
$[D_\mu,D_\nu]=-ig G_{\mu \nu}$, the twist-3 operators
$ \cO^{5 (q)}_{ [  \sigma \{ \mu_1 ] \mu_2 \cdots \mu_n \} } $
can be rewritten in a manifestly interaction-dependent form. 
For $n=2$ one finds
\begin{equation} \label{eq:twist3alt}
\cO^{5 (q)}_{ [  \sigma \{ \mu_1 ] \mu_2 \} } =
- \frac{g}{6} \bar{\psi}_q \left( \tilde{G}_{\sigma \mu_1} \gamma_{\mu_2}
    + \tilde{G}_{\sigma \mu_2} \gamma_{\mu_1} \right) \psi_q
  - \mbox{traces} \,,
\end{equation}
where $\tilde{G}^{\mu \nu}
  =\tfrac{1}{2} \epsilon^{\mu \nu \rho \sigma} G_{\rho \sigma}$ is the 
dual gluon field strength tensor and the totally antisymmetric
$\epsilon$ tensor is such that $\epsilon^{0123} = 1$.
Therefore we can define the reduced matrix element $d_2$ in the chiral limit
also by (see, e.g., Ref.~\cite{Ehrnsperger:1993hh})
\begin{equation} 
\begin{split}
&- \frac{g}{6} \langle p,s| \bar{\psi}_q
  \left( \tilde{G}_{\sigma \mu_1} \gamma_{\mu_2}
    + \tilde{G}_{\sigma \mu_2} \gamma_{\mu_1} \right) \psi_q - \mbox{traces}
  | p,s \rangle \\
&= \frac{1}{3} d_2^{(q)} [ (s_\sigma p_{\mu_1} - s_{\mu_1} p_\sigma) p_{\mu_2}
 + \cdots - \mbox{traces}] \,.
\end{split}
\end{equation}

The Wilson coefficients (\ref{eq:wilsoncoef}) can be computed in
perturbation theory, while the nucleon matrix elements $a_n^{(q)}$
and $d_n^{(q)}$ are nonperturbative quantities. For simplicity, in the
following we omit the flavor indices, in most cases.

\begin{widetext}
The renormalization of the operators which contribute to the moments of
$g_2$ has been studied by several authors in continuum perturbation 
theory~\cite{Shuryak:1981pi,Bukhvostov:1983eob,Bukhvostov:1984rns,
  Ratcliffe:1985mp,Balitsky:1987bk,Ji:1990br,Kodaira:1994ge,Kodaira:1996md}.
For example, in Refs.~\cite{Kodaira:1994ge,Kodaira:1996md} the following
operators are considered for $n=2$ in the flavor-nonsinglet sector:
\begin{align} 
 R_F^{\sigma \mu \nu} &= - \frac{i^2}{3} \left[ 
 2 \bar{\psi} \gamma^\sigma \gamma_5 D^{\{ \mu} D^{\nu \}} \psi
 - \bar{\psi} \gamma^\mu \gamma_5 D^{\{ \sigma} D^{\nu \}} \psi
 - \bar{\psi} \gamma^\nu \gamma_5 D^{\{ \mu} D^{\sigma \}} \psi \right]
 - \mbox{traces} \,, \label{eq:RF}
\\ 
 R_1^{\sigma \mu \nu} &= \frac{1}{12} g \left[ 
\epsilon^{\sigma \mu \alpha \beta}
 \bar{\psi} G_{\alpha \beta} \gamma^{\nu} \psi
+ \epsilon^{\sigma \nu \alpha \beta}
 \bar{\psi} G_{\alpha \beta} \gamma^{\mu} \psi \right] - \mbox{traces} \,,
 \label{eq:R1}
 \\ 
 R_m^{\sigma \mu \nu} &= - i  m
 \bar{\psi} \gamma^\sigma \gamma_5 D^{\{ \mu} \gamma^{\nu \}} \psi
 - \mbox{traces} \,, \label{eq:Rm}
\\ 
 R_{\mathrm {eq}}^{\sigma \mu \nu} &=  - \frac{i}{3}  \left[ 
 \bar{\psi} \gamma^\sigma \gamma_5 D^{\{ \mu} \gamma^{\nu \}} 
  (i \slashed{D} - m) \psi
 +  \bar{\psi} (i \slashed{D} - m) 
    \gamma^\sigma \gamma_5 D^{\{ \mu} \gamma^{\nu \}} \psi \right]
 - \mbox{traces} \,. \label{eq:Req}
\end{align} 
The gluon field strength tensor $G_{\alpha \beta}$ could alternatively
be expressed in terms of a commutator of two covariant derivatives. 
As Eqs.~\eqref{eq:twist3} and \eqref{eq:locop} show,
the matrix element $d_2$ corresponds to
the nucleon matrix elements of the renormalized operators \eqref{eq:RF}.
The operators \eqref{eq:RF}--\eqref{eq:Req} are linearly dependent:
\begin{equation} 
 R_F^{\sigma \mu \nu} = \frac{2}{3} R_m^{\sigma \mu \nu} +  
 R_1^{\sigma \mu \nu} +  R_{\mathrm {eq}}^{\sigma \mu \nu} \,.
\end{equation}
In the massless case, this relation leads to Eq.~\eqref{eq:twist3alt} 
upon application of the equations of motion.

Calculating the quark-quark-gluon three-point functions with a single
insertion of each of these operators in one-loop perturbation theory,
one sees that also a gauge-variant operator has to be taken into account
in the process of renormalization:
\begin{equation}
 R_{\mathrm {eq1}}^{\sigma \mu \nu} = - \frac{i}{3}  \left[ 
 \bar{\psi} \gamma^\sigma \gamma_5 \partial^{\{ \mu} \gamma^{\nu \}} 
  (i \slashed{D} - m) \psi
 +  \bar{\psi} (i \slashed{D} - m) 
    \gamma^\sigma \gamma_5 \partial^{\{ \mu} \gamma^{\nu \}} \psi \right]
 - \mbox{traces} \,.
\end{equation}
\end{widetext}
Of course, in physical matrix elements neither $R_{\mathrm {eq}}$ nor
$R_{\mathrm {eq1}}$ will contribute. They show up, however, in off-shell
vertex functions and influence the renormalization factors.

\section{Lattice Operators And Renormalization}
\label{sec:reno}

In the following, we use Euclidean notation. For our lattice evaluation
of the reduced matrix elements $a_2$ and $d_2$ we construct discretized
versions of the relevant operators. In the process of the renormalization
of these operators, operator mixing requires
particular attention because the discrete symmetry group $H(4)$ of a
hypercubic lattice is less restrictive than
the continuous symmetry group $O(4)$ of Euclidean spacetime.

In the case of the twist-2 matrix element $a_2$ we use the four-dimensional
multiplet of operators spanned by
\begin{equation} \label{eq:a2ops}
  \cO^5_{\{234\}} \, , \, \cO^5_{\{134\}} \, , \,
  \cO^5_{\{124\}} \, , \, \cO^5_{\{123\}} \,,
\end{equation}
where
\begin{equation}
\cO^5_{\sigma \mu \nu} =
  \bar{\psi} \gamma_\sigma \gamma_5 \Dlr_\mu \Dlr_\nu \psi \,.
\end{equation}
The operators \eqref{eq:a2ops} transform according to the representation
$\tau^{(4)}_3$ of the hypercubic group
$H(4)$~\cite{Baake:1981qe,Mandula:1982us,Gockeler:1996mu}. They have mass
dimension five and charge conjugation parity $+1$. These properties ensure
that they do not mix with any other gauge-invariant operators of the same
or lower dimension. Therefore, they are multiplicatively renormalizable.
We take the corresponding renormalization factor from Table XI of
Ref.~\cite{Bali:2020lwx}.

For the evaluation of the twist-3 matrix element $d_2$ we use multiplets
of operators with charge conjugation parity $+1$ which transform under
the hypercubic group according to the representation $\tau^{(8)}_1$.
Among the gauge-invariant operators of dimension $\leq 5$ there are three
multiplets that have these symmetry properties and can therefore mix
with each other under renormalization. Suitable bases transforming according
to the same (not just equivalent) unitary representation of $H(4)$ are
\begin{equation} \label{eq:d2op5}
\begin{split}
\cO^{(1)}_1 &= \frac{1}{4 \sqrt{3}}
  \left( 2 \cO^5_{2 \{14\}} - \cO^5_{1 \{24\}} - \cO^5_{4 \{12\}} \right) \,, \\
\cO^{(2)}_1 &= \frac{1}{4 \sqrt{3}}
  \left( 2 \cO^5_{2 \{13\}} - \cO^5_{1 \{23\}} - \cO^5_{3 \{12\}} \right) \,, \\
\cO^{(3)}_1 &= \frac{1}{4 \sqrt{3}}
  \left( 2 \cO^5_{3 \{14\}} - \cO^5_{1 \{34\}} - \cO^5_{4 \{13\}} \right) \,, \\
\cO^{(4)}_1 &= \frac{1}{4 \sqrt{3}}
  \left( 2 \cO^5_{3 \{24\}} - \cO^5_{2 \{34\}} - \cO^5_{4 \{23\}} \right) \,, \\
\cO^{(5)}_1 &= \frac{1}{4}
  \left( \cO^5_{4 \{23\}} - \cO^5_{2 \{34\}} \right) \,, \\
\cO^{(6)}_1 &= \frac{1}{4}
  \left( \cO^5_{4 \{13\}} - \cO^5_{1 \{34\}} \right) \,, \\
\cO^{(7)}_1 &= \frac{1}{4}
  \left( \cO^5_{4 \{12\}} - \cO^5_{1 \{24\}} \right) \,, \\
\cO^{(8)}_1 &= \frac{1}{4}
  \left( \cO^5_{3 \{12\}} - \cO^5_{1 \{23\}} \right) \,,
\end{split}
\end{equation}

\begin{equation} \label{eq:d2opsigma}
\begin{split}
\cO^{(1)}_2 &= \frac{1}{2 \sqrt{2}} \bar{\psi} 
  \left( \gamma_3 \gamma_1 \Dlr_1 - \gamma_3 \gamma_4 \Dlr_4 \right) \,, \\ 
\cO^{(2)}_2 &= \frac{1}{2 \sqrt{2}} \bar{\psi} 
  \left( \gamma_4 \gamma_3 \Dlr_3 - \gamma_4 \gamma_1 \Dlr_1 \right) \,, \\ 
\cO^{(3)}_2 &= \frac{1}{2 \sqrt{2}} \bar{\psi} 
  \left( \gamma_2 \gamma_4 \Dlr_4 - \gamma_2 \gamma_1 \Dlr_1 \right) \,, \\ 
\cO^{(4)}_2 &= \frac{1}{2 \sqrt{2}} \bar{\psi} 
  \left( \gamma_1 \gamma_2 \Dlr_2 - \gamma_1 \gamma_4 \Dlr_4 \right) \,, \\ 
\cO^{(5)}_2 &= \frac{1}{2 \sqrt{6}} \bar{\psi} 
  \left(  2 \gamma_1 \gamma_3 \Dlr_3 - \gamma_1 \gamma_2 \Dlr_2
          - \gamma_1 \gamma_4 \Dlr_4 \right) \psi \,, \\
\cO^{(6)}_2 &= \frac{1}{2 \sqrt{6}} \bar{\psi} 
  \left( -2 \gamma_2 \gamma_3 \Dlr_3 + \gamma_2 \gamma_1 \Dlr_1
          + \gamma_2 \gamma_4 \Dlr_4 \right) \psi \,, \\
\cO^{(7)}_2 &= \frac{1}{2 \sqrt{6}} \bar{\psi} 
  \left(  2 \gamma_3 \gamma_2 \Dlr_2 - \gamma_3 \gamma_1 \Dlr_1
          - \gamma_3 \gamma_4 \Dlr_4 \right) \psi \,, \\
\cO^{(8)}_2 &= \frac{1}{2 \sqrt{6}} \bar{\psi} 
  \left( -2 \gamma_4 \gamma_2 \Dlr_2 + \gamma_4 \gamma_1 \Dlr_1
          + \gamma_4 \gamma_3 \Dlr_3 \right) \psi \,,
\end{split}
\end{equation}

\begin{equation} \label{eq:d2op0}
\begin{split}
\cO^{(1)}_3 &= \bar{\psi} \left(
  \gamma_1 \Dlr_{[3} \Dlr_{1]} - \gamma_4 \Dlr_{[3} \Dlr_{4]} \right) \,, \\
\cO^{(2)}_3 &= \bar{\psi} \left(
  \gamma_3 \Dlr_{[4} \Dlr_{3]} - \gamma_1 \Dlr_{[4} \Dlr_{1]} \right) \,, \\
\cO^{(3)}_3 &= \bar{\psi} \left(
  \gamma_4 \Dlr_{[2} \Dlr_{4]} - \gamma_1 \Dlr_{[2} \Dlr_{1]} \right) \,, \\
\cO^{(4)}_3 &= \bar{\psi} \left(
  \gamma_2 \Dlr_{[1} \Dlr_{2]} - \gamma_4 \Dlr_{[1} \Dlr_{4]} \right) \,, \\
\cO^{(5)}_3 &= \frac{1}{\sqrt{3}} \bar{\psi} 
  \left(   2 \gamma_3 \Dlr_{[1} \Dlr_{3]} - \gamma_2 \Dlr_{[1} \Dlr_{2]}
           - \gamma_4 \Dlr_{[1} \Dlr_{4]} \right) \psi \,, \\
\cO^{(6)}_3 &= \frac{1}{\sqrt{3}} \bar{\psi} 
  \left( - 2 \gamma_3 \Dlr_{[2} \Dlr_{3]} + \gamma_1 \Dlr_{[2} \Dlr_{1]}
           + \gamma_4 \Dlr_{[2} \Dlr_{4]} \right) \psi \,, \\
\cO^{(7)}_3 &= \frac{1}{\sqrt{3}} \bar{\psi} 
  \left(   2 \gamma_2 \Dlr_{[3} \Dlr_{2]} - \gamma_1 \Dlr_{[3} \Dlr_{1]}
           - \gamma_4 \Dlr_{[3} \Dlr_{4]} \right) \psi \,, \\
\cO^{(8)}_3 &= \frac{1}{\sqrt{3}} \bar{\psi} 
  \left( - 2 \gamma_2 \Dlr_{[4} \Dlr_{2]} + \gamma_1 \Dlr_{[4} \Dlr_{1]}
           + \gamma_3 \Dlr_{[4} \Dlr_{3]} \right) \psi \,.
\end{split}
\end{equation}

The lattice operators \eqref{eq:d2op5} are Euclidean counterparts of the
Minkowski operators \eqref{eq:RF}, while the operators \eqref{eq:d2op0}
correspond to the operators \eqref{eq:R1} with the field strength
expressed in terms of a commutator of two covariant derivatives.
The operators
\eqref{eq:d2opsigma} are analogous to \eqref{eq:Rm}. Under renormalization
all three multiplets are expected to mix with each other. In the continuum,
$R_m$ disappears in the chiral limit. On the lattice, the explicit breaking
of chiral symmetry caused by Wilson-type fermions persists even for
massless quarks. Therefore $\cO^{(i)}_2$ will contribute with a coefficient 
$\propto a^{-1}$, where $a$ denotes the lattice spacing. The operators
$\cO^{(i)}_3$, on the contrary, are of the same dimension as $\cO^{(i)}_1$
and mix with a coefficient of order $g^2$, which should be small. The
same holds for lattice counterparts of $R_{\mathrm {eq}}$ and
$R_{\mathrm {eq1}}$. Hence, in a first approximation we take into account
only the multiplets \eqref{eq:d2op5} and \eqref{eq:d2opsigma}.
Their renormalization and mixing can be treated along the lines of
Ref.~\cite{Bali:2020lwx}, provided one multiplies the operators
\eqref{eq:d2opsigma} with $1/a$. Both operator multiplets then have
dimension five.

The renormalized operators of the multiplet \eqref{eq:d2op5} are now
given by
\begin{equation} \label{eq:mix}
\cO^{(i) \mathrm R}_1 = \hat{Z}_{11}(\mu,a) \cO^{(i)}_1 +
            \frac{1}{a} \hat{Z}_{12}(\mu,a) \cO^{(i)}_2 \,.
\end{equation}
Here we stick to the notation of Ref.~\cite{Bali:2020lwx}, where
$\hat{Z}_{m m'}(\mu,a)$ denotes the renormalization and mixing matrix in the
nonperturbative scheme used on the lattice. For this scheme we choose
the RI$^\prime$-MOM scheme, i.e., the operators are taken at
vanishing momentum. In order to suppress powerlike lattice artifacts
as far as possible the external quark momenta are chosen as
\begin{equation}
\frac{\mu}{2} (1,1,1,1)
\end{equation}
with the renormalization scale $\mu$. Presently we cannot convert the
coefficients $\hat{Z}_{m m'}(\mu,a)$ (and hence the renormalized operators) to
the $\overline{\mathrm{MS}}$ scheme, because the required perturbative
calculations in the continuum are not yet available. Our procedure
accounts for the mixing with lower-dimensional operators caused by the
explicit breakdown of chiral symmetry in our simulations, but further
mixing effects are still neglected.

In the chiral limit, the matrix element $d_2$ is multiplicatively
renormalizable~\cite{Shuryak:1981pi}. Rewriting Eq.~\eqref{eq:mix} as
\begin{equation}
\cO^{(i) \mathrm R}_1 = \hat{Z}_{11}(\mu,a) \left( \cO^{(i)}_1 +
    \frac{1}{a} \frac{\hat{Z}_{12}(\mu,a)}{\hat{Z}_{11}(\mu,a)}
    \cO^{(i)}_2 \right) \,,
\end{equation}
we see that $\cO^{(i) \mathrm R}_1$ will have a multiplicative dependence
on $\mu$ if the ratio $\hat{Z}_{12}(\mu,a)/\hat{Z}_{11}(\mu,a)$ does not
depend on $\mu$. It turns out that this requirement is better fulfilled
when we use instead of the $2 \times 2$ matrix $\hat{Z}(\mu,a)$ the
matrix $\tilde{Z}(\mu,a)$ constructed in the following way,
cf.\ Ref.~\cite{Arthur:2010ht}. We compute
$\hat{Z}(\mu,a) \hat{Z}^{-1}(\mu_0,a)$ for the renormalization scales
$\mu$ of interest and a reference scale $\mu_0$ chosen as
$\mu_0 = 2 \, \mathrm{GeV}$.
Within our approximations, this matrix should have a continuum limit,
which we evaluate by fitting the lattice spacing dependence with a
quadratic polynomial in $a^2$. Denoting the result $R(\mu,\mu_0)$,
we define
\begin{equation}
\tilde{Z}(\mu,a) = R(\mu,\mu_0) \hat{Z}(\mu_0,a) \,.
\end{equation}

To improve on this would require the consideration of quark-quark-gluon matrix
elements instead of quark-quark matrix elements.

\section{Simulation details}
\subsection{Lattice setup}\label{subsec:LatticeSetup}
To compute the reduced matrix elements $a_n^{(f)}(\mu)$ and $d_n^{(f)}(\mu)$
in~\eqref{eq:twist2} and~\eqref{eq:twist3} for $n=2$ we analyze a subset
of the lattice gauge ensembles generated within the Coordinated Lattice
Simulations (CLS) effort~\cite{Bruno:2014jqa}. The ensembles have been
generated using a tree-level Symanzik improved gauge action with
$N_f = 2+1$ flavors of nonperturbatively $O(a)$ improved Wilson (clover)
fermions. Near zero modes of the Wilson-Dirac operator are avoided
by applying twisted-mass determinant reweighting to achieve stable
Monte Carlo sampling~\cite{Luscher:2012av}. Furthermore we improve the
overlap of the interpolating currents at the source/sink time slice
using Wuppertal smeared quarks~\cite{Gusken:1989qx} in the source/sink
interpolators with APE smoothed spatial gauge links~\cite{Falcioni:1984ei}.

\begin{table*}
\caption{\label{tab:CLS}CLS gauge ensembles analyzed in this work. The
ensemble identifier is given in the first column, followed by the inverse
gauge coupling $\beta$ and the lattice size. The column `bc' indicates
whether the boundary condition in time was open (o) or (anti)periodic (p).
In the next columns we give the lattice spacing $a$, the pion mass $m_\pi$
and the product of the spatial lattice size $L = a N_s$ with the pion mass.
The column `$t/a$' contains the list of source-sink distances analyzed
on this lattice. The subscript \#meas specifies how many measurements
have been performed for the respective source-sink distance. In physical
units these distances roughly correspond to 0.9~fm, 1.0~fm and 1.2~fm.
The number of analyzed configurations is given in the column
`$n_{\rm{cnfgs}}$', and `Traj.' specifies the trajectory in the quark
mass plane to which the ensemble belongs.}
\begin{ruledtabular}
\begin{tabular}{cclcccccrc}
 Ens. & $\beta$ & \multicolumn{1}{c}{$N_s^3 \times N_t $} & bc & $a$[fm] & $m_{\pi}$ [MeV] & $L m_{\pi}$ & $t/a_{\text{\#meas}}$ & \multicolumn{1}{c}{$n_{\rm{cnfgs}}$} & Traj. \\ \hline
          A654 & 3.34 & $24^3 \times 48$ & p & 0.0984 & 334 & 4.0 & $9_4$, $11_4$, $13_4$ & 2534 & trm\\
          A653 & 3.34 & $24^3 \times 48$ & p & 0.0984 & 426 & 5.1 & $9_4$, $11_4$, $13_4$ & 2525 & trm, symm\\
          H106 & 3.4 & $32^3 \times 96$ & o & 0.0859 & 272 & 3.8 & $10_2$, $12_3$, $14_4$ & 1544 & msc\\
          H105 & 3.4 & $32^3 \times 96$ & o & 0.0859 & 279 & 3.9 & $10_2$, $12_3$, $14_4$ & 2065 & trm\\
          H102 & 3.4 & $32^3 \times 96$ & o & 0.0859 & 352 & 4.9 & $10_2$, $12_3$, $14_4$ & 2005 & trm\\
          H107 & 3.4 & $32^3 \times 96$ & o & 0.0859 & 366 & 5.1 & $10_2$, $12_3$, $14_4$ & 1561 & msc\\
          H101 & 3.4 & $32^3 \times 96$ & o & 0.0859 & 420 & 5.9 & $10_2$, $12_2$, $14_2$ & 2016 & trm, symm\\
          D451 & 3.46 & $64^3 \times 128$ & p & 0.0760 & 217 & 5.4 & $11_4$, $13_4$, $16_4$ & 531 & msc\\
          N450 & 3.46 & $48^3 \times 128$ & p & 0.0760 & 285 & 5.3 & $11_4$, $13_4$, $16_4$ & 1129 & msc\\
          B452 & 3.46 & $32^3 \times 64$ & p & 0.0760 & 350 & 4.3 & $11_4$, $13_4$, $16_4$ & 1941 & msc\\
          S400 & 3.46 & $32^3 \times 128$ & o & 0.0760 & 352 & 4.3 & $11_4$, $13_4$, $16_4$ & 2872 & trm\\
          B450 & 3.46 & $32^3 \times 64$ & p & 0.0760 & 418 & 5.2 & $11_4$, $13_4$, $16_4$ & 1612 & trm, symm\\
          N201 & 3.55 & $48^3 \times 128$ & o & 0.0643 & 285 & 4.5 & $14_2$, $16_3$, $19_4$ & 1520 & msc\\
          N203 & 3.55 & $48^3 \times 128$ & o & 0.0643 & 345 & 5.4 & $14_2$, $16_3$, $19_4$ & 1543 & trm\\
          N204 & 3.55 & $48^3 \times 128$ & o & 0.0643 & 351 & 5.5 & $14_2$, $16_3$, $19_4$ & 1500 & msc\\
          N202 & 3.55 & $48^3 \times 128$ & o & 0.0643 & 411 & 6.4 & $14_2$, $16_2$, $19_4$ & 899 & trm, symm\\
          J304 & 3.7 & $64^3 \times 192$ & o & 0.0497 & 260 & 4.2 & $17_3$, $21_3$, $24_4$ & 1630 & msc\\
          N302 & 3.7 & $48^3 \times 128$ & o & 0.0497 & 346 & 4.2 & $17_2$, $21_3$, $24_4$ & 2201 & trm\\
          N304 & 3.7 & $48^3 \times 128$ & o & 0.0497 & 351 & 4.3 & $17_2$, $21_3$, $24_4$ & 1652 & msc\\
          N300 & 3.7 & $48^3 \times 128$ & o & 0.0497 & 422 & 5.1 & $17_2$, $21_2$, $24_4$ & 500 & trm, symm\\
          J501 & 3.85 & $64^3 \times 192$ & o & 0.0391 & 333 & 4.2 & $22_2$, $27_3$, $32_4$ & 750 & trm
\end{tabular}
\end{ruledtabular}
\end{table*}

In most of our simulations we use open boundary conditions in time.
Especially for the very fine lattices this avoids freezing of the
topological charge and large autocorrelation
times~\cite{Luscher:2011kk,Luscher:2012av}.
In order to suppress the distortions caused by the loss of translation
invariance in time we restrict our measurements to regions with sufficiently
large distances from the temporal boundaries, see, e.g.,
Refs.~\cite{Bruno:2016plf,Bali:2020lwx}. Only a few of the coarser
lattices have been simulated using (anti)periodic boundary conditions.
An overview of the gauge ensembles used in this work is given in
Table~\ref{tab:CLS}. They have been generated along three different trajectories
in the quark mass plane, which are indicated in the last column of the table.
Along the trajectory labeled by `trm', the trace of the quark mass matrix is
held constant, approximately equal to its physical
value~\cite{Bruno:2014jqa}. Along the trajectory labeled by `msc', the
renormalized strange quark mass is set to its physical
value~\cite{Bali:2016umi}, and the symmetric line with equal masses of the
light quarks and the strange quark is labeled by `symm'.
A general explanation of this strategy can be found in~\cite{Bali:2016umi}.
In summary we use six different lattice spacings ranging from about
$0.039\,\mathrm{fm}$ up to $0.098\,\mathrm{fm}$ and $m_{\pi}$ goes
down from $\sim 420\,\mathrm{MeV}$ to $\sim 220\,\mathrm{MeV}$.
With linear spatial lattice extents $L m_{\pi} = a N_s m_{\pi}$
between 3.8 and 6.4, finite volume effects are expected to be moderate.
Removing the leading $O(a)$ discretization effects in the relevant
matrix elements, requires Symanzik improvement of the corresponding
operators. This has not been implemented in our study.

The extraction of the reduced matrix elements $a_n^{(f)}(\mu)$ and
$d_n^{(f)}(\mu)$ relies on the computation of ratios between three-
and two-point functions. The evaluation of the two-point functions
requires only the inversion of the lattice Dirac operator by means
of common numerical solvers. In particular, we use a modified
version of the Wuppertal adaptive algebraic
multigrid code DD-$\alpha$AMG~\cite{Babich:2010qb,Frommer:2013fsa} on
the Xeon Phi architecture~\cite{Heybrock:2015kpy,Richtmann:2016kcq,
Georg:2017diz,Georg:2017zua} and the IDFLS
solver~\cite{Luscher:2007se,Luscher:2007es}
on x86-64. The three-point functions are computed with the help
of the sequential source method~\cite{Martinelli:1988rr}, extensively
applying the so-called coherent sink method used by the LHPC Collaboration
in Ref.~\cite{Bratt:2010jn}. All the computations are performed using
the {\sc Chroma} software package~\cite{Edwards:2004sx} and additional libraries
implemented by our group.

\subsection{Correlation functions}

In Sec.~\ref{sec:OPE} we used the OPE to relate the moments of the structure
functions $g_i(x, Q^2)$ to the reduced matrix elements $a_n^{(f)}(\mu)$ and
$d_n^{(f)}(\mu)$ and specified their definitions in Eqs.~\eqref{eq:twist2}
and \eqref{eq:twist3}. The corresponding matrix elements are extracted on the
lattice from two- and three-point functions of the form
\begin{align}
  C_{\mathrm{2pt}}^{\vec{p}}(t) &=
  P_+^{\alpha \beta} \, C_{\mathrm{2pt}, \beta \alpha}^{\vec{p}}(t) 
  \nonumber \\
  = a^3 \,&\sum_{\vec{x}} e^{-i \, \vec{p} \cdot \vec{x}} \, P_+^{\alpha \beta}
   \langle \mathcal N^{\beta}(\vec{x}, t) \, \overline{\mathcal N}^{\alpha}(\vec{0}, 0)
       \rangle \,, \\
  C_{\mathrm{3pt}, \Gamma}^{\vec{p},\vec{p}^{\, \prime},\mathcal{O}}(t, \tau) &=
  \Gamma^{\alpha \beta} \,  C_{\mathrm{3pt}, \beta \alpha}^{\vec{p},
                    \vec{p}^{\, \prime}, \mathcal{O}}(t, \tau)  \nonumber \\
  = a^6 \,\sum_{\vec{x}, \vec{y}} &
   e^{-i \, \vec{p}^{\, \prime} \cdot \vec{x} +
       i (\vec{p}^{\, \prime} - \vec{p}) \cdot \vec{y}}
\nonumber \\ {} \times &  \Gamma^{\alpha \beta}
   \langle \mathcal N^{\beta}(\vec{x}, t) \, \mathcal{O}(\vec{y}, \tau) \,
      \overline{\mathcal N}^{\alpha}(\vec{0}, 0) \rangle \,.
\end{align}
The initial (final) momentum is denoted by $\vec{p}$
($\vec{p}^{\, \prime}$). The quantities of interest in this work
allow us to restrict the kinematics to the forward limit, thus we use
$\vec{p} = \vec{p}^{\, \prime}$ from now on. The nucleon is created
by the interpolating current $\overline{\mathcal N}$
at the source time slice $t_{\mathrm{src}}=0$ and annihilated at the sink
time slice $t$. In the case of the three-point correlation function an
additional local current $\mathcal{O}$ is inserted at the time slice $\tau$
with $t > \tau > 0$. The nucleon interpolating current is defined by
\begin{equation} \label{eq:nucleonInterpolator}
\mathcal N^{\alpha}(\vec{x}, t) =
  \left(u(\vec{x}, t)^{T} \, C \, \gamma_5 \, d(\vec{x}, t) \right)
  u^{\alpha}(\vec{x}, t) \,, 
\end{equation}
where $C$ is the charge conjugation matrix and the quark fields are smeared
separately in all spatial directions using the techniques mentioned in
Sec.~\ref{subsec:LatticeSetup}. Furthermore, we define the
positive parity projector $P_+ = (1 + \gamma_4) / 2$,
and $\Gamma = P_+ (-i) \gamma_j \gamma_5$ corresponds to the difference
between the two spin projections with respect to the direction $j = 1,2,3$.

When evaluating the three-point functions we consider quark-line connected 
diagrams only. Calculating the quark-line disconnected diagrams is 
computationally very expensive, but probably only of secondary
importance for the physical quantities such as the color Lorentz force on 
a quark in a nucleon. However, we should keep in mind that, strictly 
speaking, only flavor-nonsinglet quantities like $d_2^{(u)}-d_2^{(d)}$ 
are free of quark-line disconnected contributions.

The correlation functions are related to matrix elements by inserting
complete sets of energy eigenstates. In the limit of large Euclidean
times $t$, $\tau$ and $t - \tau$ excited states are exponentially
suppressed and the correlation functions can be approximated by the
ground-state contribution,
\begin{align}
C_{\mathrm{2pt}}^{\vec{p}}(t) &\approx \sum_{\sigma}
P_+^{\alpha \beta} \langle 0 | \mathcal N^{\beta} | N_{\sigma}^{\vec{p}} \rangle
    \, \langle N_{\sigma}^{\vec{p}} | \overline{\mathcal N}^{\alpha} | 0 \rangle
    \, \frac{e^{-E_{\vec{p}} t}}{2 \, E_{\vec{p}}} \,, \label{eq:c2pt} \\
  C_{\mathrm{3pt}, \Gamma}^{\vec{p},\mathcal{O}}(t, \tau)
    &\approx \sum_{\sigma, \sigma'} \, \Gamma^{\alpha \beta} \,
    \langle 0 | \mathcal N^{\beta} | N_{\sigma'}^{\vec{p}} \rangle \nonumber \\
    &\times \langle N_{\sigma'}^{\vec{p}} | \mathcal{O} |
                               N_{\sigma}^{\vec{p}} \rangle
    \langle N_{\sigma}^{\vec{p}} | \overline{\mathcal N}^{\alpha} | 0 \rangle \,
    \frac{e^{-E_{\vec{p}} t }}{4 E_{\vec{p}}^2} \,, \label{eq:c3pt}
\end{align}
where $|N_{\sigma}^{\vec{p}} \rangle$ denotes a nucleon state with spin
projection $\sigma$ and momentum $\vec{p}$. The overlap matrix elements
can be written as
\begin{equation}
P_+^{\alpha \beta}
  \langle 0 | \mathcal N^\beta (\vec{0}, 0) | N_{\sigma}^{\vec{p}} \rangle =
P_+^{\alpha \beta} \sqrt{Z_{\vec{p}}} \, u_{\sigma}^{\vec{p}, \beta}
\end{equation}
in terms of the momentum and smearing-dependent overlap factors
$Z_{\vec{p}}$ and the nucleon spinor $u_{\sigma}^{\vec{p}, \beta}$.
Similarly, the matrix elements of $\cO$ can be expressed in the form
\begin{equation}
\langle N_{\sigma'}^{\vec{p}} | \mathcal{O} | N_{\sigma}^{\vec{p}} \rangle
= \overline{u}_{\sigma'}^{\vec{p}} J[\cO] u_\sigma^{\vec{p}} \,.
\end{equation}

Using the spinor identity
$\sum_{\sigma} u_{\sigma}^{\vec{p}} \overline{u}_{\sigma}^{\vec{p}}
= E_{\vec{p}} \gamma_4 - i \vec{p} \cdot \vec{\gamma} + m_N$,
we rewrite \eqref{eq:c2pt} and \eqref{eq:c3pt} as
\begin{align}
C_{\mathrm{2pt}}^{\vec{p}}(t)
  &= Z_{\vec{p}} \, \frac{E_{\vec{p}} + m_N}{E_{\vec{p}}} \, e^{-E_{\vec{p}} t}
    + \cdots \label{eq:c2ptExpansion}  \,, \\
C_{\mathrm{3pt}, \Gamma}^{\vec{p}, \mathcal{O}}(t, \tau)
  &= Z_{\vec{p}} \, e^{-E_{\vec{p}} t} \, B_{\Gamma ,\mathcal{O}}^{\vec{p}}
    \, + \cdots \,, \label{eq:c3ptExpansion}
\end{align}
where
\begin{align}
B_{\Gamma ,\mathcal{O}}^{\vec{p}} = \frac{1}{4 E_{\vec{p}}^2} \, \mathrm{Tr}
\big \{ \Gamma \, &(E_{\vec{p}} \gamma_4 - i \vec{p} \cdot \vec{\gamma} + m_N)
J[\cO] \nonumber \\ {} \times
&(E_{\vec{p}} \gamma_4 - i \vec{p} \cdot \vec{\gamma} + m_N) \big \}.
\end{align}
The relations between the ground-state matrix elements
$\langle N_{\sigma'}^{\vec{p}} | \mathcal{O} | N_\sigma^{\vec{p}} \rangle$
and the reduced matrix elements are given in Sec.~\ref{sec:OPE}.
However, in addition to the ground-state contributions we have to take
into account possible excited states in Eqs.~\eqref{eq:c2ptExpansion}
and \eqref{eq:c3ptExpansion}. An analysis of the first excited-state
contribution is given in the next subsection.

\subsection{Excited-state contributions}

In the two- and three-point functions \eqref{eq:c2ptExpansion} and
\eqref{eq:c3ptExpansion}, respectively, the signal-to-noise ratio
decreases exponentially with the source-sink separation in time.
However, for small time distances between the operators we still
find significant excited-state contributions.
We take these contributions into account by including excited-state terms
in the spectral decomposition of the correlation
functions. Our ansatz reads
\begin{align}
  C_{\mathrm{2pt}}^{\vec{p}}(t)
  &\approx Z_{\vec{p}} \, \frac{E_{\vec{p}} + m_N}{E_{\vec{p}}} \,
  e^{-E_{\vec{p}} t} \left( 1 + A \, e^{-\Delta E_{\vec{p}} t} \right)
    \,, \label{eq:c2ptExcitedStates} \\
  C_{\mathrm{3pt}, \Gamma}^{\vec{p},\mathcal{O}}(t, \tau)
  &\approx Z_{\vec{p}} \, e^{-E_{\vec{p}} t} \,
    B_{\Gamma ,\mathcal{O}}^{\vec{p}} \, \Big( 1 \, + \,
    B_{10} e^{-\Delta E_{\vec{p}} (t - \tau)} \nonumber \\
  + &B_{01} e^{-\Delta E_{\vec{p}} \tau}
  + B_{11} e^{-\Delta E_{\vec{p}} t} \Big) \,. \label{eq:c3ptExcitedStates}
\end{align}
Here $\Delta E_{\vec{p}}$ denotes the energy difference between the first
excited state and the ground state, which is taken to be the same
in both correlators. The amplitudes of the excited-state contributions
in the two- and three-point functions are denoted by $A$ and $B$,
respectively. All amplitudes depend on the smearing and on the momentum
of the interpolating currents at the source and the sink, while
$B_{10}$, $B_{01}$ and $B_{11}$ also depend on the inserted current
$\mathcal{O}$ and the spin matrix $\Gamma$. Since we
only consider the forward limit, we may set $B_{10} = B_{01}$.

\subsection{Ratios}

\begin{figure*}
\centering
\includegraphics[scale=1.0]{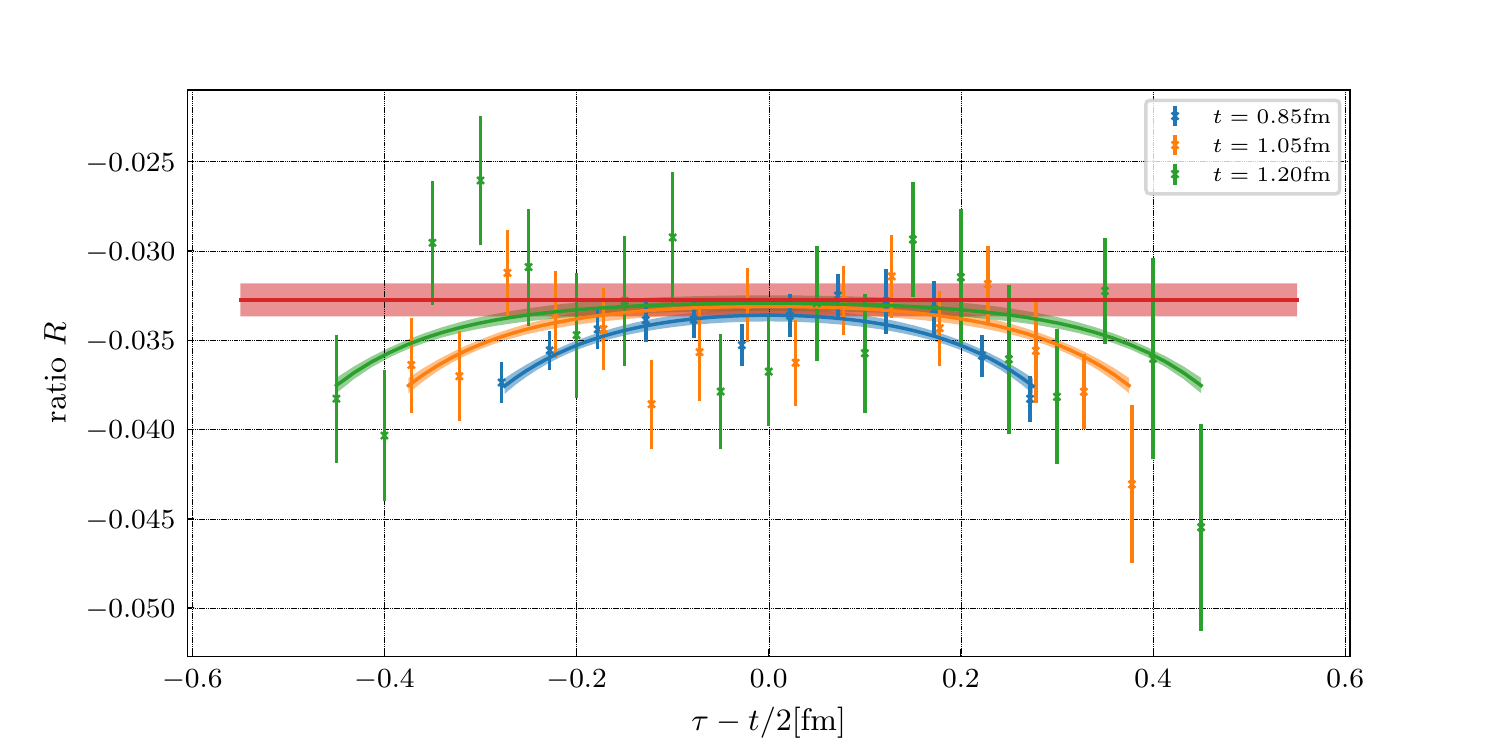}
\caption{The ratio $R$ of Eq.~\eqref{eq:ratio} for the bare operators
\eqref{eq:d2op5} with flavor $u$ on the ensemble J304 along with our fit.
The horizontal line and the corresponding error band represent the
contribution of the ground state.}
\label{fig:ratio}
\end{figure*}

Instead of performing a simultaneous fit to the two- and three-point functions,
we consider the two-point functions along with
ratios of three-point functions divided by two-point functions:
\begin{equation} \label{eq:ratio}
R_{\mathcal{O}, \Gamma}^{\vec{p}} =
\frac{C_{\mathrm{3pt}, \Gamma}^{\vec{p},\mathcal{O}}(t, \tau)}
     {C_{\mathrm{2pt}}^{\vec{p}}(t)} \stackrel{t \gg \tau \gg 0}{\approx}
\frac{E_{\vec{p}}}{E_{\vec{p}} + m_N} \,B_{\Gamma ,\mathcal{O}}^{\vec{p}} \,.
\end{equation}
In such a ratio the leading-order time dependence and the overlap factors
are eliminated and the ground-state contribution corresponds
directly to the matrix element we are interested in. Taking into account
excited-state contributions according to Eqs.~\eqref{eq:c2ptExcitedStates}
and \eqref{eq:c3ptExcitedStates}, we would arrive at the fit ansatz 
\begin{equation}
\begin{split}
R_{\mathcal{O}, \Gamma}^{\vec{p}} &\approx
\frac{E_{\vec{p}}}{E_{\vec{p}} + m_N} \,B_{\Gamma ,\mathcal{O}}^{\vec{p}}
\\ {} \times 
&\frac{1 \, + \, B_{10} e^{-\Delta E_{\vec{p}} (t - \tau)} + 
B_{01} e^{-\Delta E_{\vec{p}} \tau} + B_{11} e^{-\Delta E_{\vec{p}} t}} 
     {1 + A \, e^{-\Delta E_{\vec{p}} t}} \,.
\end{split}
\end{equation}
We assume that the ground-state energies are well described by the
continuum dispersion relation
\begin{equation}
  E_{\vec{p}} = \sqrt{\vec{p}^{\, 2} + m_N^2}
\end{equation}
in our fitting analysis. This need not be the case for $\Delta E_{\vec{p}}$
since in general multihadron states may contribute, e.g., $N \pi$ and
$N \pi \pi$ states.

\begin{table*}
\caption{Results for $d_2$ at $\mu = 2 \, \mathrm{GeV}$ obtained with the
renormalization $\tilde{Z}$ and the extrapolation functions
\eqref{eq:fit1} -- \eqref{eq:fit3}. Superscripts $(u)$ and $(d)$
refer to the $u$ and $d$ quarks in the proton.}
\label{tab:d2tilde}
\begin{ruledtabular}
\begin{tabular}{D{.}{.}{2.1} c|D{.}{.}{1.7} D{.}{.}{2.8} | ll| D{.}{.}{1.8} D{.}{.}{2.8} }
   \multicolumn{1}{c}{$\mu'^2 [\mathrm{GeV}^2]$} & Fit form &  \multicolumn{1}{c}{$d_2^{(u)}(\mu)$} & \multicolumn{1}{c|}{$d_2^{(d)}(\mu)$} & $d_2^{(u-d)}(\mu)$ & $d_2^{(u+d)}(\mu)$ & \multicolumn{1}{c}{$d_2^{(p)}(\mu)$} & \multicolumn{1}{c}{$d_2^{(n)}(\mu)$} \\ \hline
   4.0 & $f_1$ & 0.026(4) & -0.0086(26) & 0.034(4) & 0.018(5) & 0.0105(19) & -0.0009(14) \\
   8.0 & $f_1$ & 0.024(4) & -0.0090(29) & 0.033(4) & 0.016(6) & 0.0098(21) & -0.0012(15) \\
   12.0 & $f_1$ & 0.024(5) & -0.0092(31) & 0.033(5) & 0.015(6) & 0.0095(22) & -0.0013(16) \\
   4.0 & $f_2$ & 0.028(14) & 0.006(9) & 0.023(14) & 0.036(19) & 0.013(7) & 0.006(5) \\
   4.0 & $f_3$ & 0.039(13) & 0.001(9) & 0.036(13) & 0.040(18) & 0.017(6) & 0.005(5) 
\end{tabular}
\end{ruledtabular}
\end{table*}

Unfortunately, our data do not allow us to determine $B_{11}$. Therefore we
omit this term as well as the analogous contribution $\propto A$ in the
denominator from our fit function for the ratio
$R_{\mathcal{O}, \Gamma}^{\vec{p}}$. However, when performing a
simultaneous fit to the ratio \eqref{eq:ratio} and the two-point function 
\eqref{eq:c2ptExcitedStates} to extract the reduced matrix elements,
the excited-state contribution $\propto A$ is taken into account in the
two-point function. To fix the nucleon mass in the fits we include
additional two-point correlators for $\vec{p} = \vec{0}$.
The fit range of $\tau$ is restricted to the interval $2a < \tau < t - 2a$
resulting in reasonable values of $\chi^2 / \mathrm{dof}$. We choose
on-axis momenta $\vec{p}= \pm \vec{e}_i 2 \pi/L$ with $i$ taken to be
different from the polarization direction of the nucleon, which is
determined by $\Gamma$. The final analysis utilizes the data for all
available momenta, nucleon polarizations and source-sink distances.

As an example we show in Fig.~\ref{fig:ratio} the ratio \eqref{eq:ratio}
for the bare operators \eqref{eq:d2op5} (averaged over all members of the
multiplet) with flavor $u$ on the ensemble J304. The curves represent our
fit, the horizontal line and the corresponding error band show the result
for the ground-state contribution.

\section{Results}

We present our results for the light flavors $u$ and $d$ separately, where
quark-line disconnected contributions have been neglected. However, we
consider it to be unlikely that the latter would modify our numbers
beyond the size of the other uncertainties. Superscripts $(u)$ and
$(d)$ always refer to the $u$ and $d$ quarks in the proton, while
$(p)$ and $(n)$ denote the matrix elements \eqref{eq:twist3nucleon}
for the proton and the neutron, respectively,
\begin{align}
  d_2^{(p)} &= \left(\frac{2}{3}\right)^2 d_2^{(u)}
              + \left(-\frac{1}{3}\right)^2 d_2^{(d)} \,, \\
  d_2^{(n)} &= \left(-\frac{1}{3}\right)^2 d_2^{(u)}
              + \left(\frac{2}{3}\right)^2 d_2^{(d)} \,.
\end{align}
Alternatively, one can write
\begin{align}
  d_2^{(p)} &= \frac{3}{18}d_2^{(u-d)} + \frac{5}{18}d_2^{(u+d)} \,, \\
  d_2^{(n)} &= -\frac{3}{18} d_2^{(u-d)} + \frac{5}{18}d_2^{(u+d)} \,,
\end{align}
where $d_2^{(u-d)}$ does not suffer from the omission of disconnected
diagrams. 

Approximate renormalization in the RI$^\prime$-MOM scheme, as described
in Sec.~\ref{sec:reno}, is performed at some scale $\mu'$. We attempt
a combined continuum and chiral extrapolation using the fit functions
\eqref{eq:fit1} -- \eqref{eq:fit3} given below and evolve the results
to the scale $\mu = 2 \, \mathrm{GeV}$ with the help of the one-loop
formula for the flavor-nonsinglet operators,
\begin{equation} \label{eq:d2run}
d_2(\mu) = \left(\frac{\alpha_s(\mu')}{\alpha_s(\mu)}\right)^{-B}d_2(\mu') \,,
\end{equation}
where
\begin{equation} \label{eq:d2anodim}
B = \frac{1}{\frac{11}{3}N_c - \frac{2}{3} N_f}
  \left(3 N_c - \frac{1}{6}\left(N_c - \frac{1}{N_c}\right)\right)
\end{equation}
with $N_c=3$ and $N_f=3$. As we neglect disconnected contributions, we use
this value of $B$, which is strictly speaking only correct for the $(u-d)$
part, also for $(u+d)$. Varying the intermediate scale $\mu'$ should give
us some measure of the uncertainty related to the renormalization. The
central results, however, are all obtained at $\mu'=\mu$, so they do not
depend on the perturbative value of $B$.

When constructing our fit functions we take into account that the
leading discretization effects in the matrix elements in our
simulations are $O(a)$. For the continuum and chiral extrapolation
of the $d_2$ and $a_2$ data, we consider the fit formulas
\begin{align} \label{eq:fit1}
f_1(a,m_\pi, m_K) &= C_1 + C_2a + C_3 \delta m^2 + C_4 \bar{m}^2 \,,
\\ \label{eq:fit2}
f_2(a,m_\pi, m_K) &= C_1 + C_2a + C_3 \delta m^2 + C_4 \bar{m}^2 + C_5 a^2 \,,
\\ 
f_3(a,m_\pi, m_K) &= C_1 + C_2a + C_3 \delta m^2 + C_4 \bar{m}^2
 \nonumber  \\ \label{eq:fit3}
& \hspace*{1.93cm} + C_5 \bar{m}^2 a + C_6 \delta m^2 a \,,
\end{align}
where
\begin{equation}
\delta m^2 = m_K^2 - m_\pi^2  \quad , \quad
\bar{m}^2 = (2m_K^2 + m_\pi^2)/3 \,.
\end{equation}
The difference between the corresponding results should provide an impression
of the uncertainty inherent in the fit procedure. The gauge field
ensembles used in the fits are collected in Table~\ref{tab:CLS}.
Note that on a few configurations in some of the coarser ensembles
(H105, H106, H107, N450, B452 and N201) we encountered measurements
that were separated by more than a hundred 68\% confidence level
intervals from the results obtained on the remaining configurations.
The origin of these deviations is unclear and we excluded these
configurations from further analysis.

\begin{figure*} 
\centering
\includegraphics[scale=1.0]{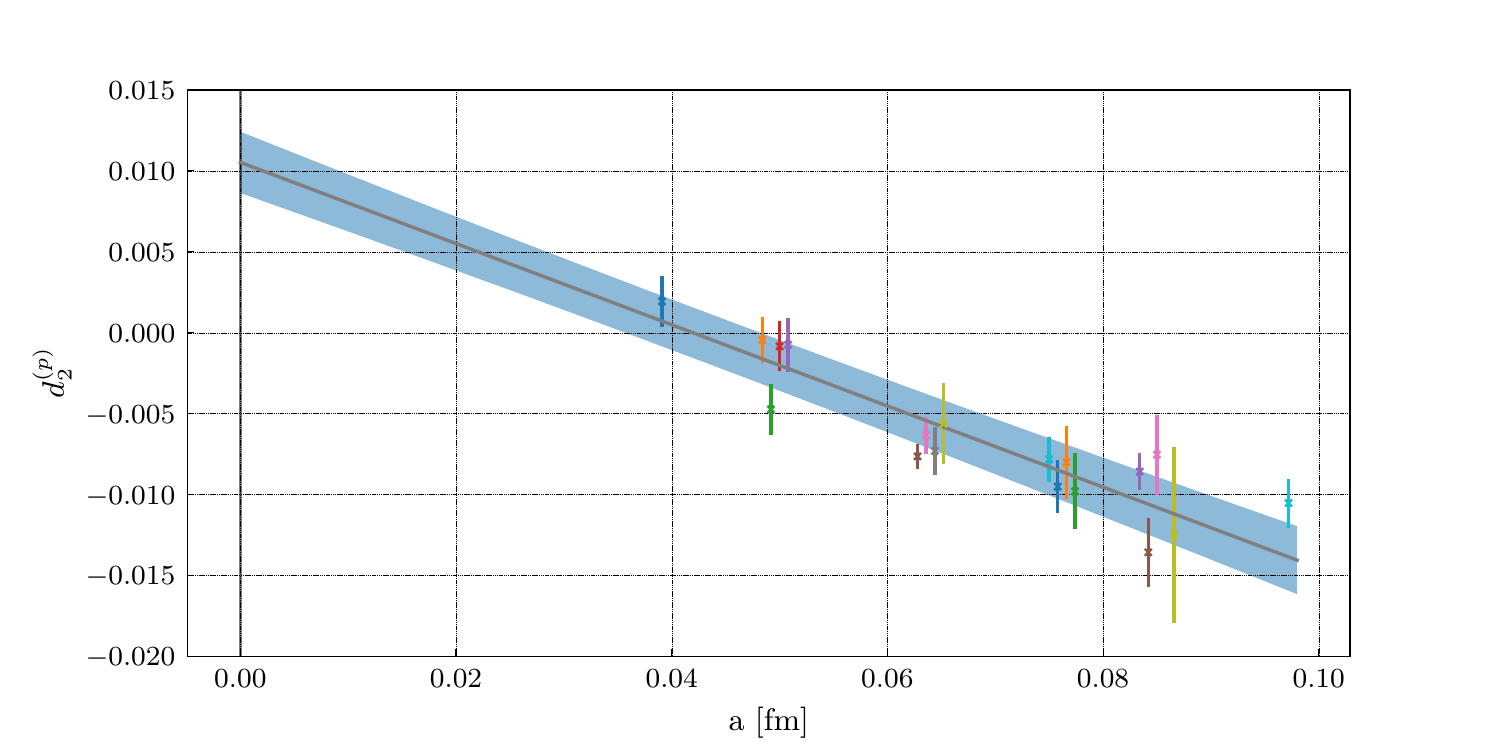}
\includegraphics[scale=1.0]{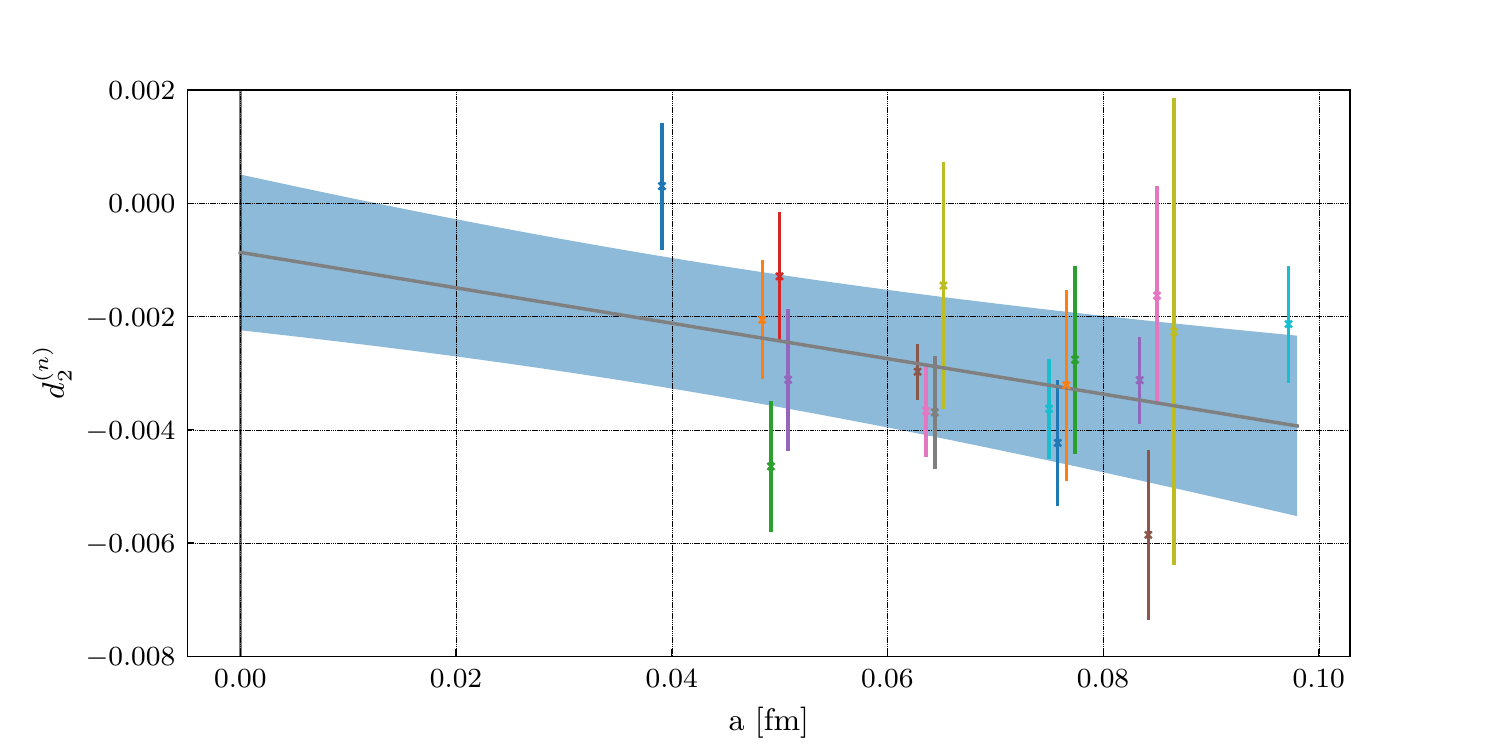}
\caption{Continuum extrapolation of $d_2^{(p)}$ and $d_2^{(n)}$.
This extrapolation corresponds to the first line of Table~\ref{tab:d2tilde}.
Results belonging to the same value of $a$ are horizontally shifted
for visibility by
a small amount such that the exact value of $a$ lies in the center of the
respective group. Within each of these groups the pion mass decreases from
left to right. The results from the ensembles A654, H105 and D451 are
not shown on the plots because of their large error bars.}
\label{fig:d2}
\end{figure*}

\begin{figure}
\centering
\includegraphics[scale=1.5]{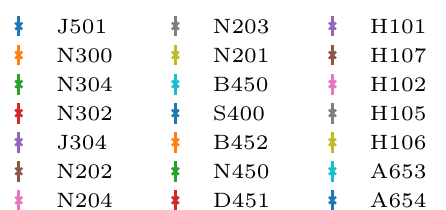}
\caption{Legend for Figs.~\ref{fig:d2}, \ref{fig:d2mass}
and \ref{fig:a2}. The order of the
ensembles (from top left to bottom right) is the same as in these figures
(from left to right).}
\label{fig:legend}
\end{figure}

The fit results for $d_2$ are given in Table~\ref{tab:d2tilde} for the
case of the renormalization matrix $\tilde{Z}$, cf.\ Sec.~\ref{sec:reno}.
The dependence on the intermediate scale $\mu'$ is quite weak,
due to the use of $\tilde{Z}$. However, the dependence on the choice
of the fit function is more pronounced, although all three functions
yield reasonable fits with $\chi^2/\mathrm{dof}$ between 0.66 and 0.94.
In Fig.~\ref{fig:d2} we plot our data along with the fit function
\eqref{eq:fit1} for $\mu' = 2 \, \mathrm{GeV}$ versus the lattice
spacing $a$. The legend identifying the ensembles is given in
Fig.~\ref{fig:legend}. The curve shows the fit function evaluated with
the physical values of $m_\pi$ and $m_K$. The fitted coefficients
$C_3$ and $C_4$ have been used to shift the data
points vertically such that they correspond to the physical masses. 

\begin{figure*} 
\centering
\includegraphics[scale=1.0]{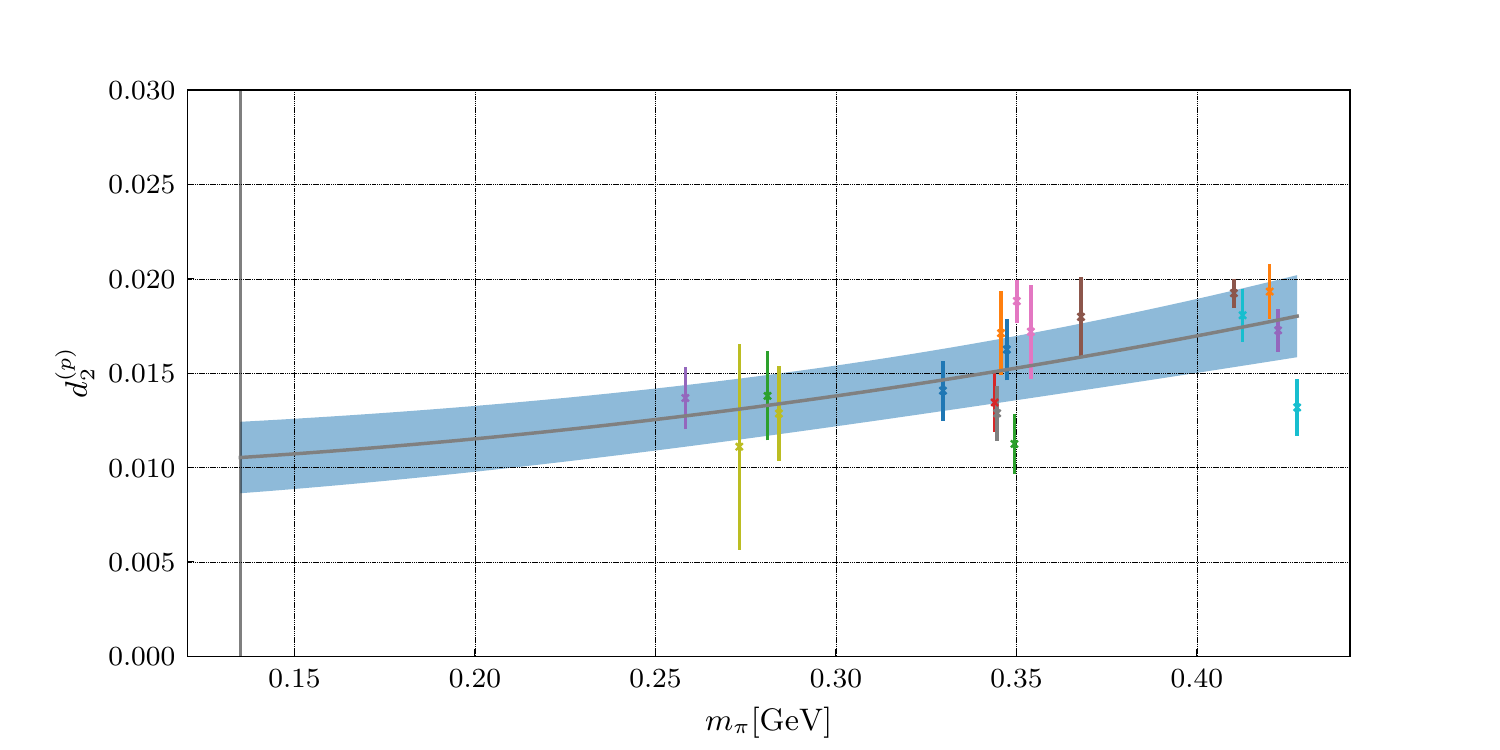}
\caption{Mass dependence of $d_2^{(p)}$.
The fit curve corresponds to the first line of Table~\ref{tab:d2tilde}.
The results from the ensembles A654, H105 and D451 are
not shown on the plot because of their large error bars.}
\label{fig:d2mass}
\end{figure*}

In order to visualize the mass dependence we show in Fig.~\ref{fig:d2mass}
the results for $d_2^{(p)}$ together with the fit corresponding to the
first line of Table~\ref{tab:d2tilde} plotted against $m_\pi$. The curve
represents the function $f_1(0,m_\pi,m_K^{\mathrm {phys}})$ evaluated
with the fitted parameters, and the data points have been shifted by
subtracting $f_1(a,m_\pi, m_K) - f_1(0,m_\pi,m_K^{\mathrm {phys}})$.
Compared to the $a$ dependence, the dependence on $m_\pi$ turns
out to be more moderate.

\begin{table*}
\caption{Results for $a_2$ in the $\overline{\mathrm{MS}}$ scheme
  at $\mu = 2 \, \mathrm{GeV}$ obtained with the extrapolation functions
  \eqref{eq:fit1} -- \eqref{eq:fit3}. Superscripts $(u)$ and $(d)$
  refer to the $u$ and $d$ quarks in the proton.}
\label{tab:a2}
\begin{ruledtabular}
\begin{tabular}{ c  c| D{.}{.}{1.7} D{.}{.}{2.7} | ll|D{.}{.}{1.7} D{.}{.}{1.7} }
  $\mu'^2 [\mathrm{GeV}^2]$ & Fit form &  \multicolumn{1}{c}{$a_2^{(u)}(\mu)$} & \multicolumn{1}{c|}{$a_2^{(d)}(\mu)$} & $a_2^{(u-d)}(\mu)$ & $a_2^{(u+d)}(\mu)$ & \multicolumn{1}{c}{$a_2^{(p)}(\mu)$} & \multicolumn{1}{c}{$a_2^{(n)}(\mu)$} \\ \hline
   4.0 & $f_1$ & 0.161(11) & -0.025(6) & 0.187(10) & 0.136(14) & 0.069(5) & 0.0068(33) \\
   4.0 & $f_2$ & 0.195(35) & -0.016(21) & 0.208(35) & 0.18(5) & 0.085(17) & 0.015(11) \\
   4.0 & $f_3$ & 0.137(33) & -0.038(19) & 0.175(31) & 0.10(4) & 0.057(15) & -0.001(10) 
\end{tabular}
\end{ruledtabular}
\end{table*}

\begin{figure*} 
\centering
\includegraphics[scale=1.0]{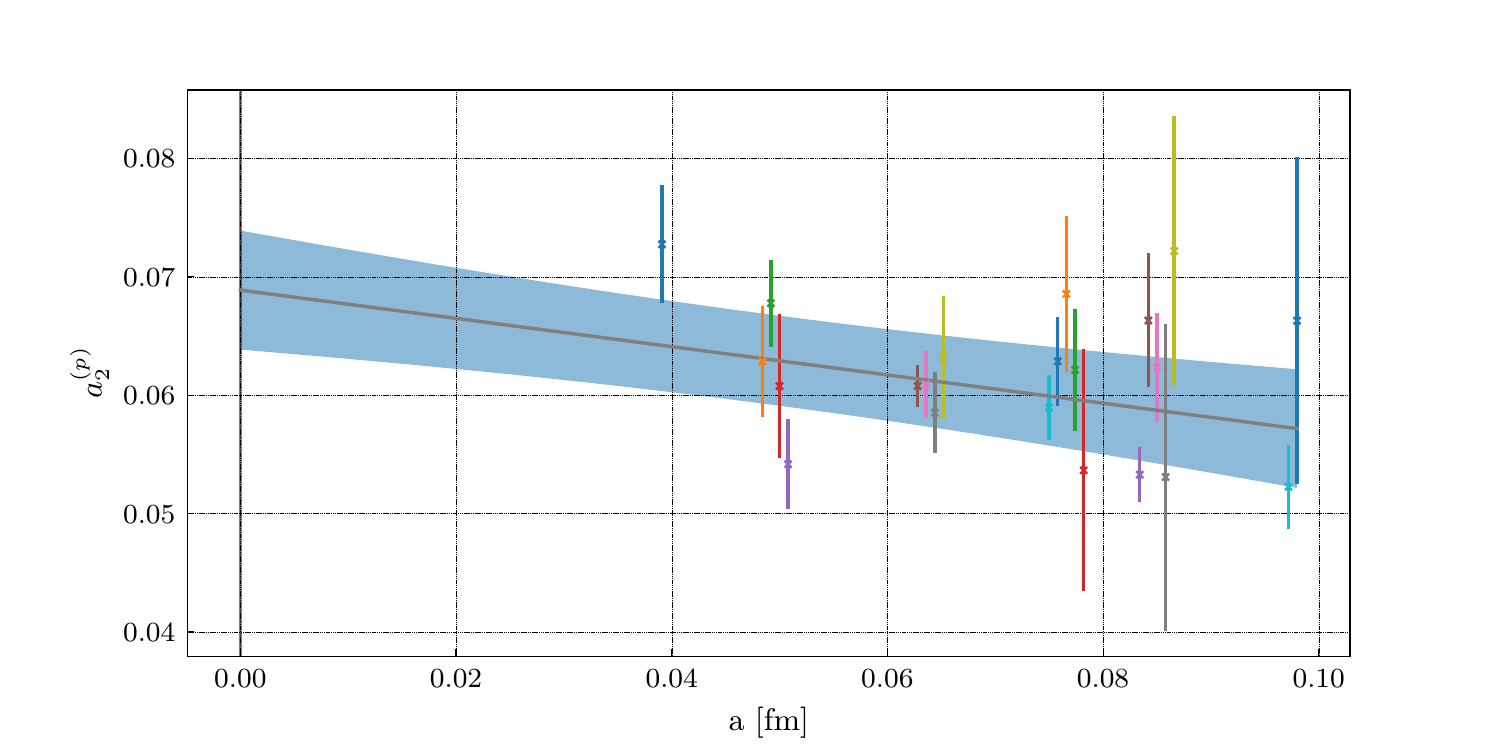}
\includegraphics[scale=1.0]{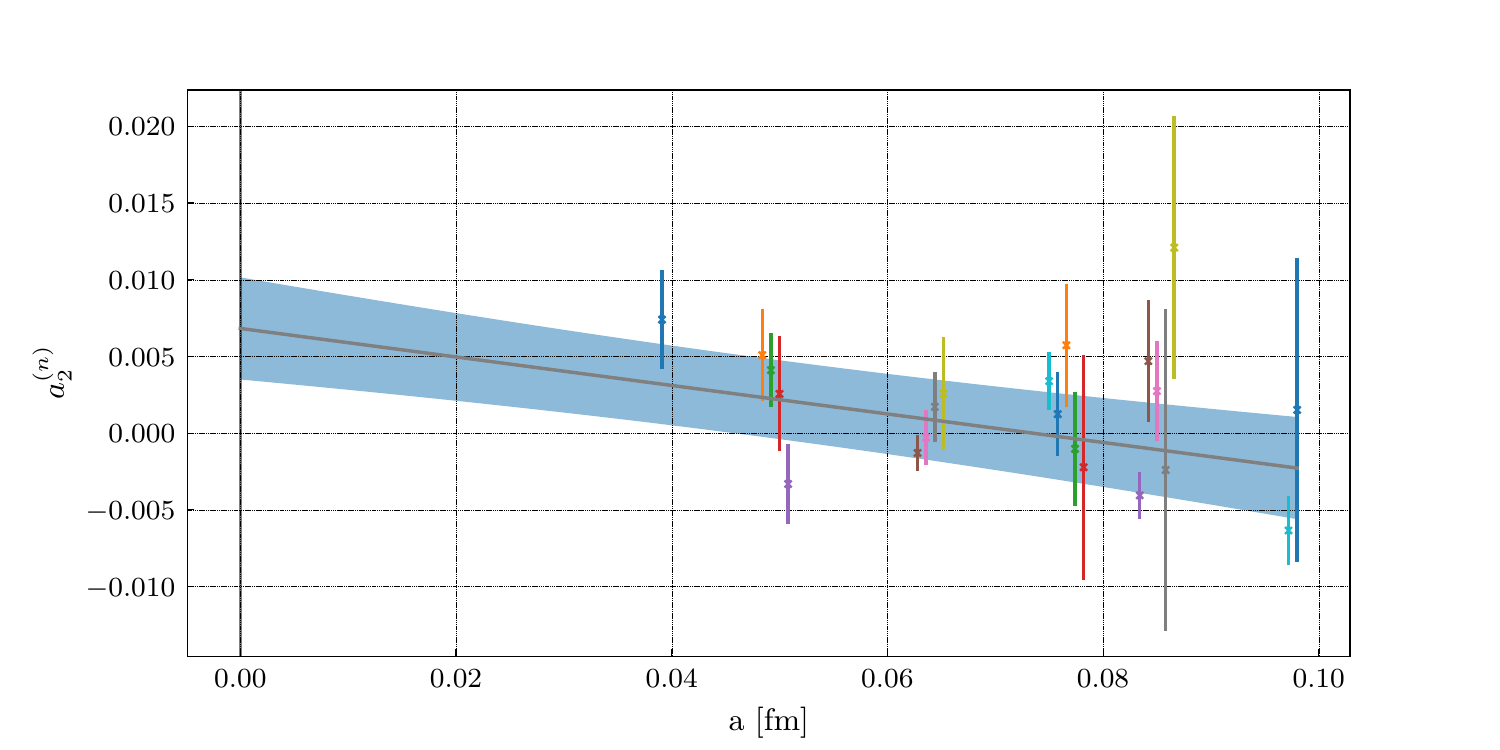}
\caption{Continuum extrapolation of $a_2^{(p)}$ and $a_2^{(n)}$.
This extrapolation corresponds to the first line of Table~\ref{tab:a2}.
Results belonging to the same value of $a$ are horizontally shifted
for visibility by
a small amount such that the exact value of $a$ lies in the center of the
respective group. Within each of these groups the pion mass decreases from
left to right.}
\label{fig:a2}
\end{figure*}

Results for the twist-2 matrix element $a_2$ are presented in
Table~\ref{tab:a2}. The flavor dependence is of the same form as in the
case of $d_2$. Since the operators that we use for the determination of
$a_2$ are multiplicatively renormalizable, we can follow the standard
RI$^\prime$-MOM procedure to obtain values in the $\overline{\mathrm{MS}}$
scheme at $\mu = 2 \, \mathrm{GeV}$. As in the case of $d_2$, we obtain
reasonable fits with all three fit functions
($0.80 \leq \chi^2/\mathrm{dof} \leq 0.92$),
but find some dependence of the results for $a_2$ on the
fit function. Plots of our $a_2$ data and the fit function \eqref{eq:fit1}
are shown in Fig.~\ref{fig:a2}, which is analogous to Fig.~\ref{fig:d2}.
Again, the dependence on the pion mass appears to be rather weak.

For our final values, collected in Table~\ref{tab:final}, we take the
results given in the first line of Table~\ref{tab:d2tilde} for $d_2$
and Table~\ref{tab:a2} for $a_2$. The errors in these tables are purely
statistical, but there are several sources of systematic uncertainties.
Comparing the results obtained by varying the fit function or the
scale $\mu'$ allows us to estimate the influence of the extrapolation
method. Since the renormalization in the case of $a_2$ is less
subtle than in the case of $d_2$, we have refrained from varying the
intermediate renormalization scale $\mu'$ in the analysis for $a_2$.
As the estimate of the systematic error due to our extrapolation
we take the maximum of the (absolute value of the) difference between
the final value and the results obtained by means of the fit functions
$f_2$ and $f_3$. Unfortunately, we are not able to assess the effect of
neglecting the operators~\eqref{eq:d2op0} in the process of the
renormalization of $d_2$. This problem must be left for future
investigations. Another source of error that cannot yet be
quantified is the omission of quark-line disconnected contributions,
which leaves only $a_2^{(u-d)}$ and $d_2^{(u-d)}$ unaffected.
However, we expect that this error is small compared to the other
uncertainties. 

\begin{table*}
\caption{Final results for $a_2$ ($\overline{\mathrm{MS}}$ scheme)
and for $d_2$ (renormalization $\tilde{Z}$), both at
$\mu = 2 \, \mathrm{GeV}$. The first error is statistical, while the
second error accounts for the uncertainty due to the combined chiral and
continuum extrapolations. The error caused by the approximations in the
renormalization procedure for $d_2$ cannot be quantified.}
\label{tab:final}
\begin{ruledtabular}
\begin{tabular}{ c | D{.}{.}{1.11} D{.}{.}{1.13} | D{.}{.}{1.11} D{.}{.}{1.11} | D{.}{.}{1.12} D{.}{.}{1.12} }
   {} &  \multicolumn{1}{c}{$\dummy_2^{(u)}(\mu)$} & \multicolumn{1}{c|}{$\dummy_2^{(d)}(\mu)$} & \multicolumn{1}{c}{$\dummy_2^{(u-d)}(\mu)$} & \multicolumn{1}{c|}{$\dummy_2^{(u+d)}(\mu)$} & \multicolumn{1}{c}{$\dummy_2^{(p)}(\mu)$} & \multicolumn{1}{c}{$\dummy_2^{(n)}(\mu)$} \\ \hline
  $\dummy=d$  & 0.026(4)(13) & -0.0086(26)(146) & 0.034(4)(11) & 0.018(5)(22) & 0.0105(19)(65) & -0.0009(14)(69) \\
 $\dummy=a$ & 0.161(11)(34) & -0.025(6)(13) & 0.187(10)(21) & 0.136(14)(44) & 0.069(5)(16) & 0.0068(33)(82)
\end{tabular}
\end{ruledtabular}
\end{table*}

\section{Comparison with experiment}

Several experiments have measured the structure functions $g_i(x,Q^2)$
for certain ranges of the variables $x$ and $Q^2$ and attempted to determine
the moments 
\begin{equation}
\int_0^1 dx \, x^2 g_i(x,Q^2) 
\end{equation}
for the proton as well as for the neutron. Results are given for $Q^2$
up to $5 \, \mathrm{GeV}^2$. As far as we can see, the Wilson coefficients
are taken into account only in leading order. In this approximation the
moments of the structure functions are related to the reduced matrix
elements by
\begin{equation} \label{eq:a2def}
a_2 (\mu) = 4 \int_0^1 dx \, x^2 g_1(x,Q^2) 
\end{equation}
and
\begin{equation} \label{eq:d2def}
d_2 (\mu) = 2 \int_0^1 dx \, x^2 \left( 2 g_1(x,Q^2) + 3 g_2(x,Q^2) \right)  
\end{equation}
with $\mu^2 = Q^2$.

Let us begin with the results for $a_2$. In Table~\ref{tab:res.a2int} we
collect the values given for $\int_0^1 dx \, x^2 g_1(x,Q^2)$ in the
literature with statistical and systematic errors (if they are given
separately) added in quadrature.

\begin{table}
\caption{Experimental results for $\int_0^1 dx \, x^2 g_1(x,Q^2)$.}
\label{tab:res.a2int}
\begin{ruledtabular}
\begin{tabular}{l D{.}{.}{1.2} D{.}{.}{1.10} D{.}{.}{1.9}}
  {} & \multicolumn{1}{c}{$Q^2 [\mathrm{GeV}^2]$} & \multicolumn{1}{c}{proton}
  & \multicolumn{1}{c}{neutron} \\ \hline
  Ref.~\cite{Abe:1998wq} & 5.0 & 0.0124(10) & -0.0024(16) \\
  Ref.~\cite{Osipenko:2005nx} & 4.2 & 0.01100(83) &  \multicolumn{1}{c}{-} \\
  Ref.~\cite{Osipenko:2005nx} & 5.0 & 0.00853(175) & \multicolumn{1}{c}{-} \\
  Ref.~\cite{Flay:2016wie} & 3.21 & \multicolumn{1}{c}{-}  & 0.00086(64) \\
  Ref.~\cite{Flay:2016wie} & 4.32 & \multicolumn{1}{c}{-}  & 0.00050(65) 
\end{tabular}
\end{ruledtabular}
\end{table}

\begin{table}
\caption{Values for $a_2$ with $\mu^2 = 4 \, \mathrm{GeV}^2$ from experiment.}
\label{tab:res.a2exp}
\begin{ruledtabular}
\begin{tabular}{l D{.}{.}{1.2} D{.}{.}{1.8} D{.}{.}{1.8}}
  {} & \multicolumn{1}{c}{$Q^2 [\mathrm{GeV}^2]$} & \multicolumn{1}{c}{$a_2^{(p)}(\mu)$} & \multicolumn{1}{c}{$a_2^{(n)}(\mu)$} \\ \hline
  Ref.~\cite{Abe:1998wq} & 5.0 & 0.0497(40) & -0.0096(64) \\
  Ref.~\cite{Osipenko:2005nx} & 4.2 & 0.0427(32) &  \multicolumn{1}{c}{-} \\
  Ref.~\cite{Osipenko:2005nx} & 5.0 & 0.0342(70) &  \multicolumn{1}{c}{-} \\
  Ref.~\cite{Flay:2016wie} & 3.21 & \multicolumn{1}{c}{-}  & 0.0032(24) \\
  Ref.~\cite{Flay:2016wie} & 4.32 & \multicolumn{1}{c}{-}  & 0.0020(25)
\end{tabular}
\end{ruledtabular}
\end{table}

In order to compare the results given in Table~\ref{tab:res.a2int} with our
numbers we take into account the one-loop QCD corrections, which are
flavor independent, i.e., we divide the experimental values of
$\int_0^1 dx \, x^2 g_1(x,Q^2)$ by the Wilson coefficient
\eqref{eq:wilsoncoef1} with $\mu^2 = Q^2$. Then we use the renormalization
group to evolve the renormalization scale of the resulting matrix element
$a_2$ to our value $\mu^2 = 4 \, \mathrm{GeV}^2$. We employ the five-loop
anomalous dimension of the relevant operator multiplet along with the
five-loop $\beta$ function. The details of the calculation are the same
as in Ref.~\cite{Bali:2020lwx}. The resulting values for
$a_2(2 \, \mathrm{GeV})$ can be found in Table~\ref{tab:res.a2exp}.

In Table~\ref{tab:res.d2half} we collect the values presented for
$d_2^{\mathrm {alt}} = d_2/2$
in the literature with statistical and systematic errors (if they are given
separately) added in quadrature. As in this case Wilson coefficients
beyond tree level are not available, we just use the renormalization
group to evolve the renormalization scale to our value
$\mu^2 = 4 \, \mathrm{GeV}^2$. The corresponding factors are calculated
from Eqs.~\eqref{eq:d2run} and \eqref{eq:d2anodim}, and the resulting
values for $d_2(2 \, \mathrm{GeV})$ can be found in Table~\ref{tab:res.d2exp}.

As in the analysis of the experimental data the variation of $Q^2$
in the respective experimental setup generally was not taken into account,
applying the renormalization group running is not too well justified, but
the effect is anyhow quite small compared to the experimental
uncertainties. If, however, the sign change from negative numbers at the
smaller values of $Q^2$ to positive numbers at larger $Q^2$ is taken
seriously and interpreted as a
``nontrivial scale dependence''~\cite{Armstrong:2018xgk}, the perturbative
renormalization group would not be applicable in this range of $Q^2$.

In addition to these results from single experiments, there is also
a global analysis of polarized inclusive deep-inelastic scattering
available~\cite{Sato:2016tuz}. Unfortunately, the resulting values
for $d_2^{(p)}$ and $d_2^{(n)}$ are given at the rather low scale
$Q^2 = 1 \, \mathrm{GeV}^2$. If one nevertheless uses Eq.~\eqref{eq:d2run}
for the evolution to the scale $4 \, \mathrm{GeV}^2$, one finds
$d_2^{(p)}(2 \, \mathrm{GeV})= 0.0062(25)$ and
$d_2^{(n)}(2 \, \mathrm{GeV})= -0.0012(12)$ in broad agreement
with the individual results in Table~\ref{tab:res.d2exp}.

\begin{table}
\caption{Experimental results for $d_2^{\mathrm {alt}} = d_2/2$. In
the case of Ref.~\cite{E154:1997eyc} we have chosen the ``SLAC average''.}
\label{tab:res.d2half}
\begin{ruledtabular}
\begin{tabular}{l D{.}{.}{1.2} D{.}{.}{1.10} D{.}{.}{1.10}}
  {} & \multicolumn{1}{c}{$Q^2 [\mathrm{GeV}^2]$} & \multicolumn{1}{c}{$d_2^{(p)}/2$} & \multicolumn{1}{c}{$d_2^{(n)}/2$} \\ \hline
  Ref.~\cite{E154:1997eyc} & 3.0 & \multicolumn{1}{c}{-} & -0.010(15) \\
  Ref.~\cite{Abe:1998wq} & 5.0 & 0.0058(50) & 0.0050(210) \\
  Ref.~\cite{Anthony:2002hy} & 5.0 & 0.0032(17) & 0.0079(48) \\
  Ref.~\cite{Zheng:2004ce} & 5.0 & \multicolumn{1}{c}{-} & 0.0062(28) \\
  Ref.~\cite{Osipenko:2005nx} & 4.2 & 0.0014(130) & \multicolumn{1}{c}{-} \\
  Ref.~\cite{Osipenko:2005nx} & 5.0 & 0.0035(150) & \multicolumn{1}{c}{-} \\
  Ref.~\cite{Airapetian:2011wu} & 5.0 & 0.0148(107) & \multicolumn{1}{c}{-} \\
  Ref.~\cite{Posik:2014usi} & 3.21 & \multicolumn{1}{c}{-} & -0.00421(114) \\
  Ref.~\cite{Posik:2014usi} & 4.32 & \multicolumn{1}{c}{-} & -0.00035(108) \\
  Ref.~\cite{Armstrong:2018xgk} & 2.8 & -0.00414(328) & \multicolumn{1}{c}{-} \\
  Ref.~\cite{Armstrong:2018xgk} & 4.3 & -0.00149(400) & \multicolumn{1}{c}{-}   
\end{tabular}
\end{ruledtabular}
\end{table}

\begin{table}
\caption{Results for $d_2$ with $\mu^2 = 4 \, \mathrm{GeV}^2$ from experiment.
In the case of Ref.~\cite{E154:1997eyc} we have chosen the ``SLAC
average''.}
\label{tab:res.d2exp}
\begin{ruledtabular}
\begin{tabular}{l D{.}{.}{1.2} D{.}{.}{1.9} D{.}{.}{1.9}}
  {} & \multicolumn{1}{c}{$Q^2 [\mathrm{GeV}^2]$} & \multicolumn{1}{c}{$d_2^{(p)}(\mu)$} & \multicolumn{1}{c}{$d_2^{(n)}(\mu)$} \\ \hline
  Ref.~\cite{E154:1997eyc} & 3.0 & \multicolumn{1}{c}{-} & -0.019(28) \\
  Ref.~\cite{Abe:1998wq} & 5.0 & 0.0122(106) & 0.0106(443) \\
  Ref.~\cite{Anthony:2002hy} & 5.0 & 0.0068(36) & 0.0167(101) \\
  Ref.~\cite{Zheng:2004ce} & 5.0 & \multicolumn{1}{c}{-} & 0.0131(59) \\
  Ref.~\cite{Osipenko:2005nx} & 4.2 & 0.0028(263) & \multicolumn{1}{c}{-} \\
  Ref.~\cite{Osipenko:2005nx} & 5.0 & 0.0074(317) & \multicolumn{1}{c}{-} \\
  Ref.~\cite{Airapetian:2011wu} & 5.0 & 0.0312(226) & \multicolumn{1}{c}{-} \\
  Ref.~\cite{Posik:2014usi} & 3.21 & \multicolumn{1}{c}{-} & -0.0080(22) \\
  Ref.~\cite{Posik:2014usi} & 4.32 & \multicolumn{1}{c}{-} & -0.0007(22) \\
  Ref.~\cite{Armstrong:2018xgk} & 2.8 & -0.0075(60) & \multicolumn{1}{c}{-} \\
  Ref.~\cite{Armstrong:2018xgk} & 4.3 & -0.0030(81) & \multicolumn{1}{c}{-}   
\end{tabular}
\end{ruledtabular}
\end{table}

Our results for $a_2^{(p)}$ in Table~\ref{tab:a2}
are larger than the experimental values,
but the dependence on the choice of the fit function may indicate
that this discrepancy should not be taken too seriously.
A similar tendency is observed in Fig.~20 of Ref.~\cite{Fan:2020nzz},
where moments obtained from quasidistributions are compared with
phenomenological determinations. Figure~6 of Ref.~\cite{Edwards:2006qx}
(see also Ref.~\cite{Bratt:2010jn}) seems to suggest that a more
sophisticated chiral extrapolation could diminish this discrepancy, but
better data are needed to clarify this issue.
Concerning $a_2^{(n)}$ we can hardly say more than that it must be
quite small, as also indicated by the experimental values.
  
The experimental results for $d_2^{(p)}$ collected in Table~\ref{tab:res.d2exp}
may perhaps be summarized in the statement that the data taken at
reasonably large $Q^2$ hint at a value in the vicinity of 0.01 for
$\mu = 2 \, \mathrm{GeV}$. The results for $d_2^{(n)}$ are not so easy to
summarize. While the order of magnitude is 0.01 as well, even the sign
is ambiguous. For the final numbers from our lattice calculation see
Table~\ref{tab:final}. We find a value for $d_2^{(n)}$ which is
consistent with zero. For $d_2^{(p)}$ we get a number quite close to 0.01,
which is consistent with most of the experimental determinations.

\section{Comparison with other lattice determinations}

There are a few previous lattice investigations of $a_2$ and $d_2$,
with which we can compare our new results. For this purpose we consider
the reduced matrix elements $a_2^{(q)}$ and $d_2^{(q)}$ with $q=u,d$.

In Ref.~\cite{Gockeler:2005vw} a continuum limit was not attempted.
Instead, the results on the finest lattice in the chiral limit were
considered as the best estimates obtained in this $N_f=2$ simulation.
With the help of the perturbative running factor we evolve our $a_2$
results from $\mu^2 = 4 \, \mathrm{GeV}^2$ to the scale
$\mu^2 = 5 \, \mathrm{GeV}^2$ used in Ref.~\cite{Gockeler:2005vw}.
This yields $a_2^{(u)}= 0.155(11)(33)$ and $a_2^{(d)}= -0.024(6)(13)$
to be compared with $a_2^{(u)}= 0.187(28)$ and $a_2^{(d)}= -0.056(11)$.
In the case of $d_2$ we obtain $d_2^{(u)}= 0.025(4)(12)$ and
$d_2^{(d)}= -0.0081(25)(138)$ at $\mu^2 = 5 \, \mathrm{GeV}^2$ to be
compared with the values $d_2^{(u)}= 0.010(12)$ and $d_2^{(d)}= -0.0056(50)$
given in Ref.~\cite{Gockeler:2005vw}. The statistical errors of our
present determination are significantly smaller than those quoted
in Ref.~\cite{Gockeler:2005vw}, while the central values are in
rough agreement with each other. Unfortunately, the uncertainties
due to the combined chiral and continuum extrapolations,
which could not be estimated in the previous study, are still rather large.

Values for $\langle x^2 \rangle_{\Delta u}= a_2^{(u)}/2$ and 
$\langle x^2 \rangle_{\Delta d}= a_2^{(d)}/2$ from another $N_f=2$ 
simulation are presented in Ref.~\cite{LHPC:2002xzk}. They are obtained
at a single lattice spacing $a \approx 0.1\,\mathrm{fm}$ with the
help of perturbative renormalization. Since the results are given 
at $\mu^2 = 4 \, \mathrm{GeV}^2$ in the $\overline{\mathrm{MS}}$ scheme,
they can directly be compared with our numbers. From Table~IX of 
Ref.~\cite{LHPC:2002xzk} we get  $a_2^{(u)} = 0.232(84)$ and 
$a_2^{(d)} = 0.002(50)$. Although not extrapolated to the continuum,
these values are roughly compatible with our results.

Somewhat indirect information on $a_2^{(u-d)}$ is contained in 
Ref.~\cite{Bratt:2010jn}. This paper relies on simulations with
$N_f=2+1$ flavors at a single lattice spacing $a = 0.124 \,\mathrm{fm}$
and employs a combination of perturbative and nonperturbative methods
for the renormalization. Results are given at a renormalization scale
$\mu^2 \approx 2.5 \, \mathrm{GeV}^2$ for the generalized form factors
$\tilde{A}_{10}^{u-d} (t)$ and $\tilde{A}_{30}^{u-d} (t)$, which are
related to $\langle x^2 \rangle_{\Delta u - \Delta d}= a_2^{(u-d)}/2$ 
through
\begin{equation}
\langle x^2 \rangle_{\Delta u - \Delta d} = \tilde{A}_{30}^{u-d} (0)
= \frac{\tilde{A}_{30}^{u-d} (t)}{\tilde{A}_{10}^{u-d} (t)} \Bigg |_{t=0}
\tilde{A}_{10}^{u-d} (0)
\end{equation}
with $\tilde{A}_{10}^{u-d} (0) = g_A$. Reading off from Fig.~32 in
Ref.~\cite{Bratt:2010jn} the value 
\begin{equation}
\frac{\tilde{A}_{30}^{u-d} (t)}{\tilde{A}_{10}^{u-d} (t)} \Bigg |_{t=0}
\approx 0.09
\end{equation}
and using $g_A \approx 1.27$ one obtains $a_2^{(u-d)} \approx 0.229$
for $\mu^2 \approx 2.5 \, \mathrm{GeV}^2$. This value corresponds to
$a_2^{(u-d)} \approx 0.209$ at $\mu^2 = 4 \, \mathrm{GeV}^2$,
consistent with our result in Table~\ref{tab:final}.

Results from the quasidistribution approach~\cite{Bhattacharya:2020cen}
lead to the conclusion that the Wandzura-Wilczek approximation for
$g_T(x) = g_1(x) + g_2(x)$ works well at least up to $x \approx 0.4$.
Whether the deviations observed at higher values of $x$ indicate
nonvanishing twist-3 effects or have a different origin, remains to be seen.

\section{Conclusions}\label{sec:conclusions}

In the present paper we computed the nucleon matrix elements $a_2$ and
$d_2$, which determine the $x^2$ moments of the spin structure functions
$g_1$ and $g_2$, in lattice QCD, thus improving on the earlier
evaluation~\cite{Gockeler:2005vw} (coauthored by some of us).
In both determinations quark-line disconnected contributions were neglected.

\begin{figure} 
\centering
\includegraphics[scale=0.4]{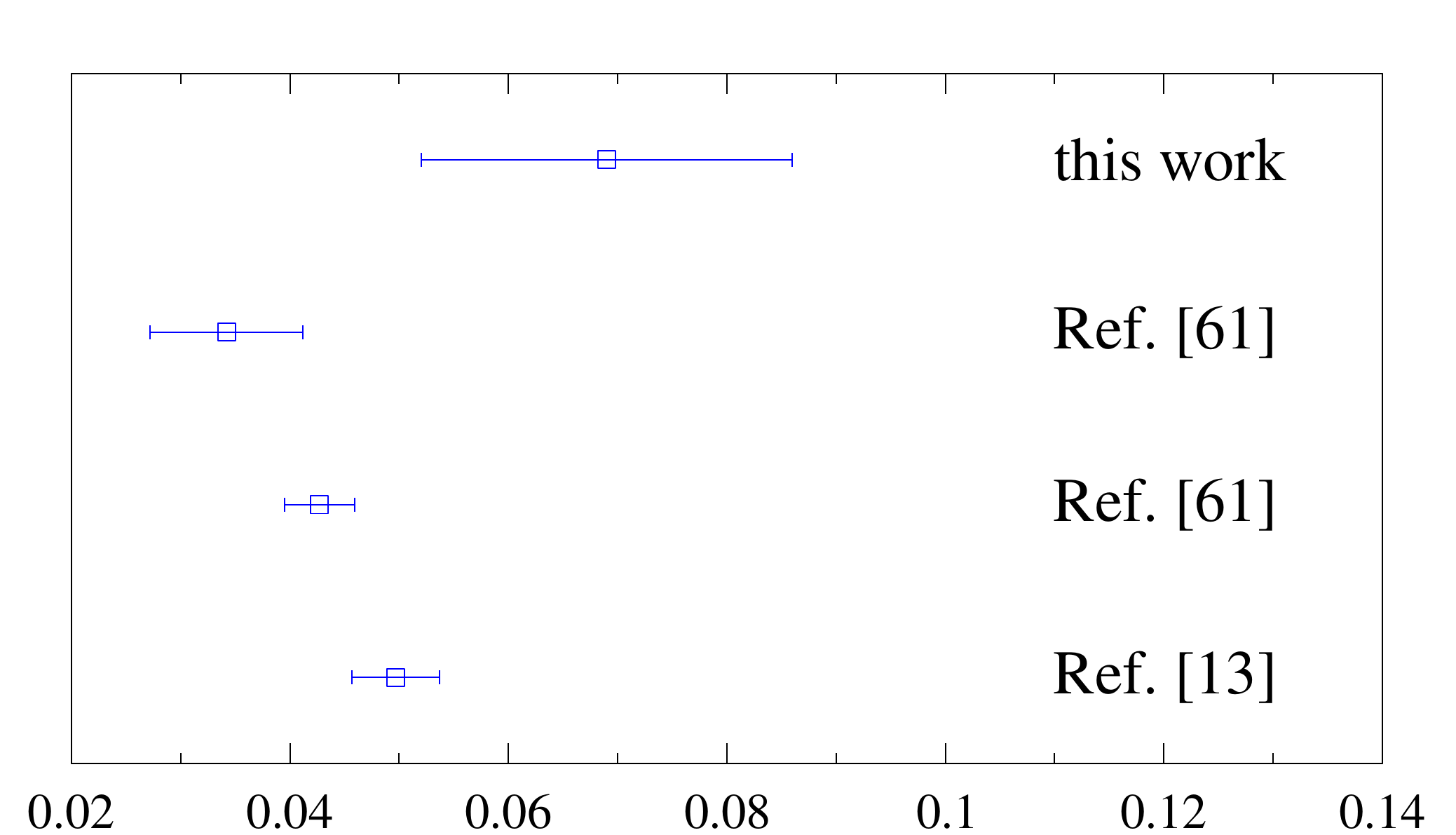}
\caption{Comparison of our results for $a_2^{(p)}$ with experimental
values. The renormalization scale is $\mu^2 = 4 \, \mathrm{GeV}^2$.
Statistical and systematic errors have been added in quadrature.}
\label{fig:comparison_a2p}
\end{figure}

\begin{figure} 
\centering
\includegraphics[scale=0.4]{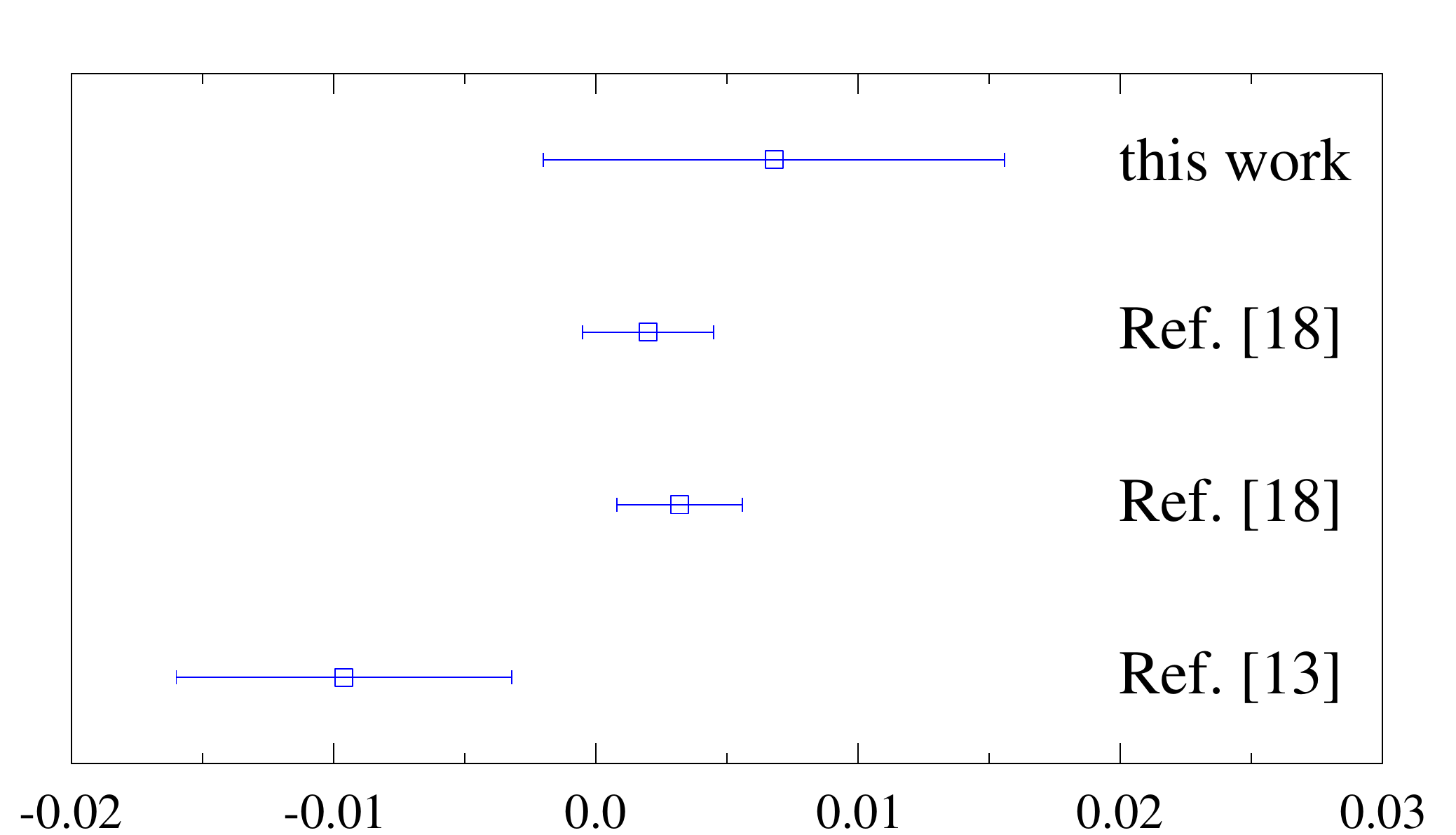}
\caption{Comparison of our results for $a_2^{(n)}$ with experimental
values. The renormalization scale is $\mu^2 = 4 \, \mathrm{GeV}^2$.
Statistical and systematic errors have been added in quadrature.}
\label{fig:comparison_a2n}
\end{figure}

\begin{figure} 
\centering
\includegraphics[scale=0.4]{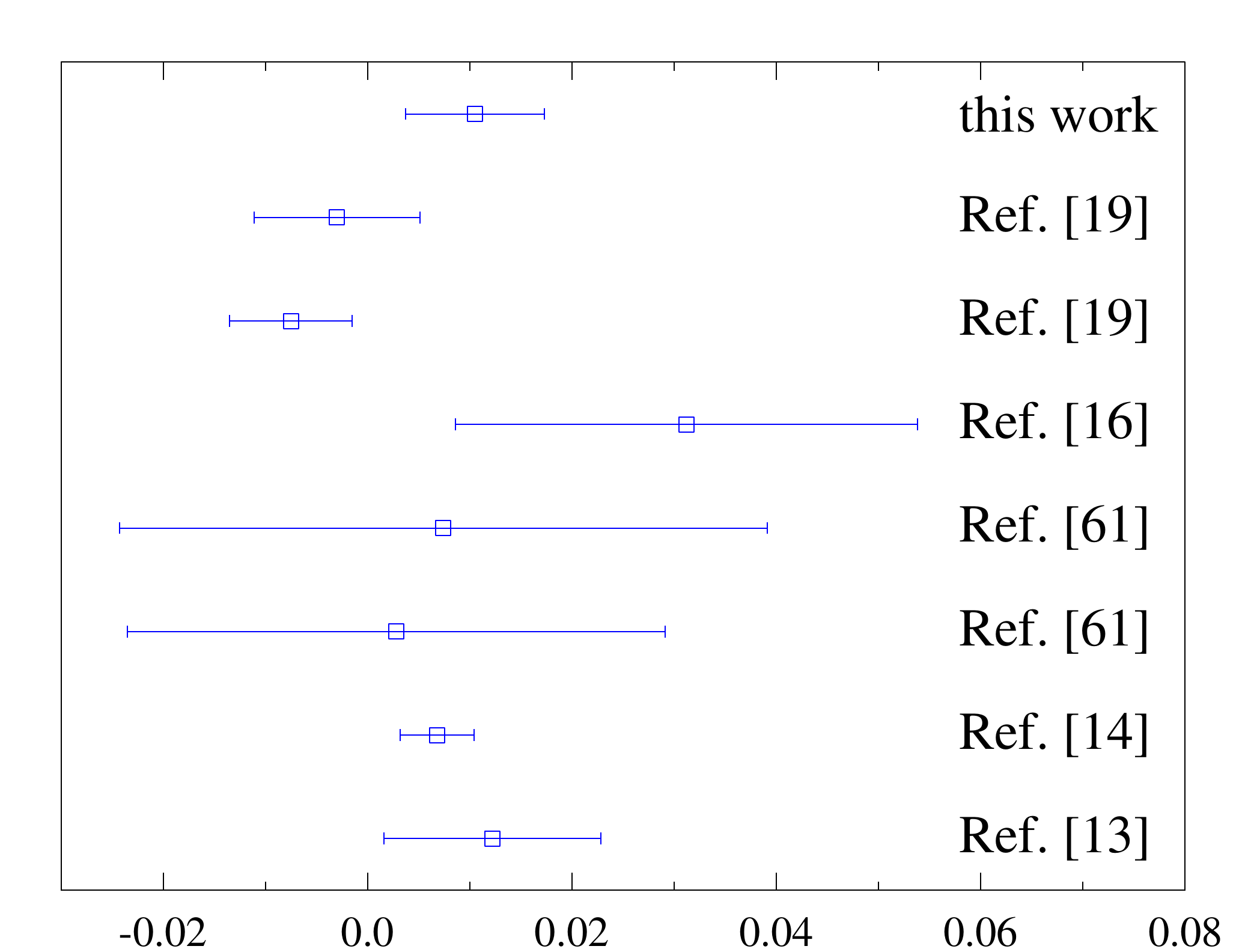}
\caption{Comparison of our results for $d_2^{(p)}$ with experimental
values. The renormalization scale is $\mu^2 = 4 \, \mathrm{GeV}^2$.
Statistical and systematic errors have been added in quadrature.}
\label{fig:comparison_d2p}
\end{figure}

\begin{figure} 
\centering
\includegraphics[scale=0.4]{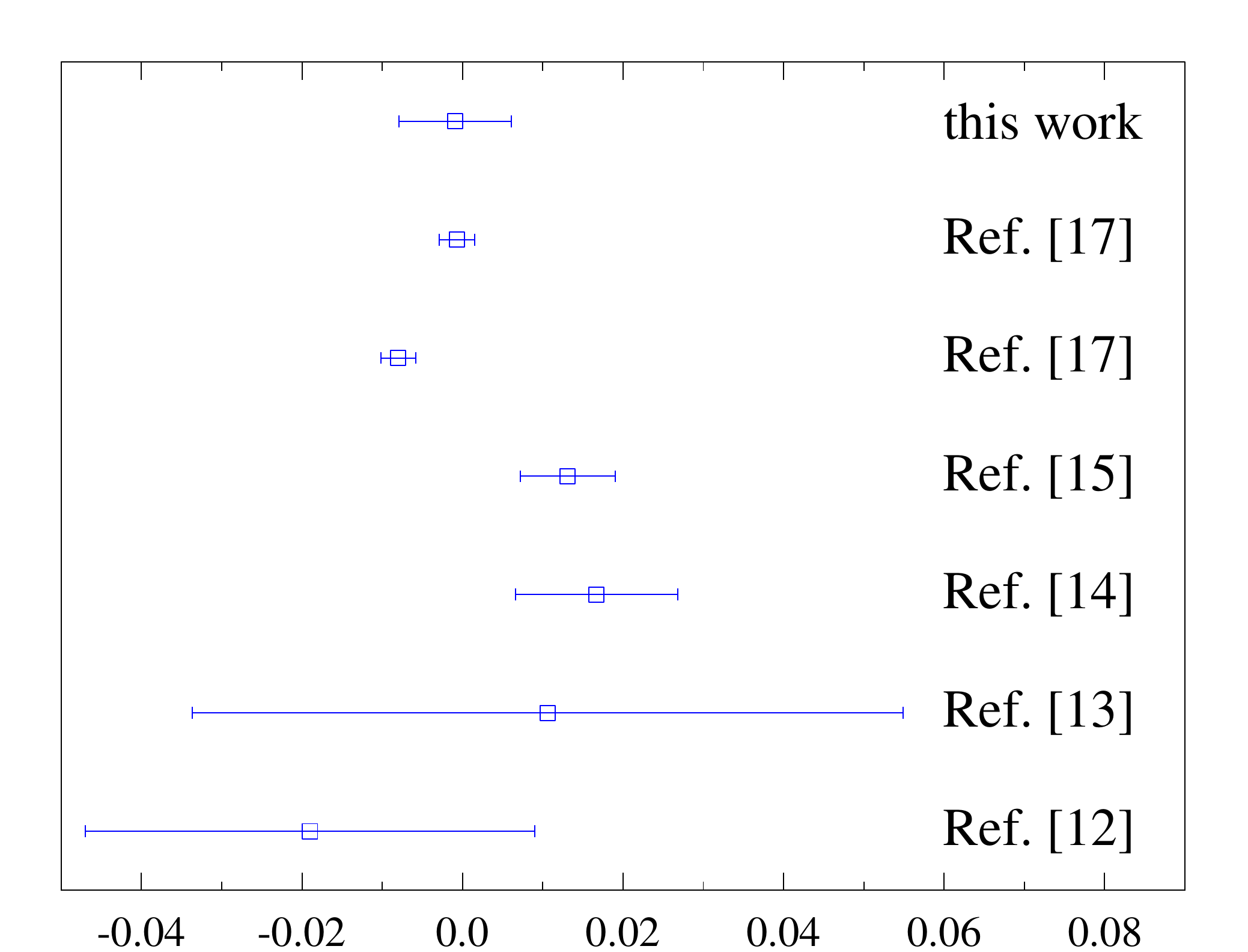}
\caption{Comparison of our results for $d_2^{(n)}$ with experimental
values. The renormalization scale is $\mu^2 = 4 \, \mathrm{GeV}^2$.
Statistical and systematic errors have been added in quadrature.}
\label{fig:comparison_d2n}
\end{figure}

For the twist-2 matrix element $a_2$ we found
$a_2^{(p)} = 0.069(5)(16)$ and
$a_2^{(n)} = 0.0068(33)(82)$ at the renormalization scale
$\mu = 2 \, \mathrm{GeV}$, see also Table~\ref{tab:final}. 
In both cases the second (systematic) error is considerably larger
than the first (statistical) error. Our results
are in broad agreement with phenomenology, as shown in
Figs.~\ref{fig:comparison_a2p} and \ref{fig:comparison_a2n}.

The parameter $d_2$ quantifies a specific twist-3 quark-gluon
correlation in the nucleon. Because it is experimentally accessible it 
became a much discussed test case for our understanding of hadron 
structure beyond twist 2. We observed a strong dependence on lattice 
spacing for $d_2^{(p)}$, see Fig.~\ref{fig:d2}, which implies that the
good agreement between the earlier lattice result and experiment
probably was somewhat accidental. In contrast, in this new
lattice determination of $d_2^{(p)}$ and $d_2^{(n)}$ we take the
lattice spacing dependence into account and only when doing
this, the results agree well with experiment (and, therefore, also
with the numbers given in Ref.~\cite{Gockeler:2005vw}). Note also
that the $a$ dependence of $d_2^{(n)}$ is far less 
pronounced. These results provide a showcase example justifying the CLS 
strategy to focus its resources on controlling the continuum limit. The 
$a$ dependence of any observable of interest can be strong (as for 
$d_2^{(p)}$) or weak (as for $d_2^{(n)}$). What is the case has to be 
carefully evaluated for each specific quantity.

The final results can be found in Table~\ref{tab:final}.
We obtained $d_2^{(p)} = 0.0105(19)(65)$ and $d_2^{(n)} = -0.0009(14)(69)$
at the renormalization scale $\mu = 2 \, \mathrm{GeV}$. Again, the
systematic error dominates the total one. In Figs.~\ref{fig:comparison_d2p}
and \ref{fig:comparison_d2n} we compare our findings with results from the
experimental and phenomenological literature.

Following Refs.~\cite{Burkardt:2008ps,Aslan:2019jis} these numbers
can be related to the transverse color Lorentz force on a quark in a
transversely polarized proton. Considering the proton in its rest
frame with $p^+ = m_N/\sqrt{2}$ and $s^x = m_N$, one gets in analogy
to Eq.~\eqref{eq:force}
\begin{equation} 
F^{q,y}(0) = m_N^2 d_2^{(q)} \,.
\end{equation}
With $m_N^2 \approx 4.47 \, \mbox{GeV/fm}$ we obtain 
\begin{equation} 
\begin{split}
F^{u,y}(0) &= 116(61) \, \mbox{MeV/fm} \,, \\
F^{d,y}(0) &= - 38(66) \, \mbox{MeV/fm} \,,
\end{split}
\end{equation} 
where we have added the two errors in quadrature.

\vspace*{0.5cm}

\section*{Acknowledgments}

The authors thank V.M.~Braun, W.~S\"oldner and S.~Weish\"aupl for
discussions and valuable input and our colleagues in the Coordinated
Lattice Simulations effort
(CLS~\cite{Bruno:2014jqa}, \url{http://wiki-zeuthen.desy.de/CLS/CLS}) for
the joint generation of the gauge field ensembles. We used a modified version
of the {\sc Chroma}~\cite{Edwards:2004sx} software package, along with
improved linear solvers~\cite{Luscher:2012av,Nobile:2010zz,Frommer:2013fsa,
Heybrock:2015kpy}. The gauge ensembles were generated as part of the
CLS effort, using {\sc OpenQCD}~\cite{LuscherOpenQCD,Luscher:2012av}.

This work was supported by the DFG (Deutsche Forschungsgemeinschaft) through
the collaborative research center SFB/TRR-55 and the Research Unit
FOR 2926 ``Next Generation pQCD for Hadron Structure: Preparing for the EIC''
and the European Union’s Horizon 2020 research and innovation program
under the Marie Sk{\l}odowska-Curie grant agreement no.~813942 (ITN EuroPLEx)
and grant agreement no.~824093 (STRONG-2020). A.~Sternbeck acknowledges
support by the BMBF (German Federal Ministry of Education and Research)
under Grant No.\ 05P15SJFAA (FAIR-APPA-SPARC) and by the DFG Research
Training Group GRK1523.

The authors gratefully acknowledge computing time granted by the
John von Neumann Institute for Computing (NIC) and provided on the
Booster partition of the supercomputer JURECA~\cite{jureca} at
J\"ulich Supercomputing Centre (JSC) as well as
the Gauss Centre for Supercomputing e.V.~(\url{https://www.gauss-centre.eu})
for providing computing time on the GCS Supercomputer
SuperMUC-NG at Leibniz Supercomputing Centre (\url{https://www.lrz.de}).
GCS is the alliance of the three national supercomputing centers
HLRS (Universit\"at Stuttgart), JSC (Forschungszentrum J\"ulich), and
LRZ (Bayerische Akademie der Wissenschaften), funded by the German
Federal Ministry of Education and Research (BMBF) and the German State
Ministries for Research of Baden-W\"urttemberg (MWK), Bayern (StMWFK) and
Nordrhein-Westfalen (MIWF). Computer time on the DFG-funded Ara cluster
at the Friedrich Schiller University Jena is acknowledged.
Additional simulations were carried out on the
Regensburg Athene2 cluster and the SFB/TRR 55 QPACE~3 machine.

\end{document}